\documentclass[aps,preprint,pra]{revtex4}

\usepackage[pdftex]{color}

\usepackage{graphicx}
\usepackage{dcolumn}
\usepackage{bm}
\usepackage{amsmath}

\newcommand{\bra}{\langle}
\newcommand{\ket}{\rangle}

\begin{document}

\title{Quantum complementarity of clocks\\ in the context of general relativity}
\author{Zhifan Zhou}
    \affiliation{Department of Physics, Ben-Gurion University of the Negev, Be'er Sheva 84105, Israel}
\author{Yair Margalit}
    \affiliation{Department of Physics, Ben-Gurion University of the Negev, Be'er Sheva 84105, Israel}
\author{Daniel Rohrlich}\email{rohrlich@bgu.ac.il}\thanks{corresponding author}
    \affiliation{Department of Physics, Ben-Gurion University of the Negev, Be'er Sheva 84105, Israel}
\author{Yonathan Japha}
    \affiliation{Department of Physics, Ben-Gurion University of the Negev, Be'er Sheva 84105, Israel}
\author{Ron Folman}
    \affiliation{Department of Physics, Ben-Gurion University of the Negev, Be'er Sheva 84105, Israel}

\begin{abstract}
Clocks play a special role at the interface of general relativity and quantum mechanics. We analyze a clock-interferometry thought experiment and go on to theoretically derive and experimentally test a complementarity relation for quantum clocks in the context of the gravitational time lag. We study this relation in detail and discuss its application to various types of quantum clocks.
\end{abstract}

\maketitle

\section{Introduction}
The interface between quantum mechanics (QM) and general relativity (GR) is an ongoing fundamental challenge.  While cosmology and high-energy physics offer tools used for probing this interface and seeking hints for a highly sought-after unification, here our tools are table-top spatial atomic interferometry and atomic clocks.  Indeed, progress in matter-wave interferometry \cite{Kasevich,Tower,muller} and atomic clocks \cite{JunYe} has provided a promising platform for new experiments. To unambiguously test the GR notion of proper time in the context of QM, a self-interfering clock has been suggested \cite{Zych,Bushev}. Such a scheme has recently been realized in a proof-of-principle experiment \cite{Us}. Quantum complementarity \cite{Bohr} plays a special role at this QM--GR interface, as we show below.

Our present understanding of complementarity \cite{Englert, Greenberg, Jaeger, Mandel, Zurek} for a two-path interferometer is summarized by the fundamental inequality $V^2+D^2 \leq 1$, where $V$ is interference pattern visibility and $D$ is distinguishability of the two paths of the interfering particle. This law has been verified in numerous experiments \cite{Rauch, Aspect, Rempe, MandelExp, Haroche2, Electron, Pfau, Chapman} and elaborated theoretically \cite{Bergou,NCinternal}. In the framework of GR, there is speculation \cite{Zych} that the inequality may be broken such that $V^2+D^2 > 1$. As clock interferometry sensitive to gravitational red shifts may soon be feasible \cite{JunYe, Wineland,Campbell, Huntemann, Chen, Ludlow}, formulating an account of clock complementarity is timely. Here we analyze in detail, and test experimentally, a clock complementarity rule for spatial interferometers with internal Hilbert spaces. See also \cite{brukner} for a closely related analysis.  We begin with a clock-interferometry thought experiment, suggesting a clock complementarity rule in the context of proper time. We obtain it theoretically for an atomic clock with two or more internal levels, and verify it empirically in a clock interferometry experiment that includes a simulated gravitational red shift.

\section{Theory}
\label{sec:theory}

In the thought experiment, a clock is prepared in a spatial superposition where one wave packet is closer to a gravitational source and thus suffers from a stronger time lag (or red shift) \cite{Zych,Us}. We note that it has been theoretically shown that spatial interferometers which are sensitive to a proper time lag between the paths are possible \cite{Wolfgang1}.  Now, on the one hand, if the ``ticking" rate of the clock depends on its path, then clock time provides which-path information and the inequality $V^2+D^2 \leq 1$, developed in the framework of non-relativistic QM, must apply.  Yet, on the other hand, gravitational time lags do not arise in non-relativistic quantum mechanics, which is not covariant and therefore not consistent with the equivalence principle \cite{Lugli}.  Hence our treatment of the clock superposition is a semiclassical extension of quantum mechanics to include gravitational red shifts.

As a historical precedent, we note that at the sixth Solvay conference in 1930, Einstein tried to defeat the uncertainty principle for time and energy by using a clock to measure the precise time a photon is released, and a spring scale to weigh the change in energy $E$ (via $E=mc^2$) of the whole apparatus.  Bohr then applied gravitational time dilation to show that Einstein's suggestion could not succeed \cite{BE}.  Indeed, Bohr's reply to Einstein already contains the idea for our thought experiment, if we transform the uncertain height of the clock in the gravitational potential (during the weighing) into a superposition of the clock at different heights.  Yet Bohr's refutation seems, at first sight, mysterious.  How could Bohr have applied something outside of quantum mechanics to refute a quantum-mechanical argument?  Isn't quantum mechanics by itself, without general relativity, a self-consistent theory?  The explanation \cite{AR} is simple:  Einstein suggested measuring the energy of a photon by $weighing$ it; he thus equated the $inertial$ mass $m$ (in the formula for energy) with the $gravitational$ mass (in the weight of the photon).  But this equation --- the equivalence principle --- implies the red shift!  In this work we reverse the logical implication: since we impose a red shift, we must also impose the equivalence principle.

According to the equivalence principle, two wave packets traversing an interferometer in a gravitational field can equivalently be described as two wave packets traversing the interferometer and accelerating \cite{feynman lectures}.  That is, we can map the experiment with its gravitational field to an equivalent experiment with no gravitational field, but with acceleration; and relativistic quantum mechanics fully describes the latter experiment.  It follows that the two experiments are equivalent; for otherwise, quantum mechanics could distinguish between them, contradicting the equivalence principle. It likewise follows that complementarity, which is expected to hold also for relativistic QM, should also apply to wave packets that acquire different red shifts.

An atomic clock accumulates a quantum phase between two or more internal levels. It is convenient to represent clock states as vectors $\bf{s}$ in the Bloch sphere. In Fig. 1(a) we show two such vectors corresponding to two interfering clock wave packets. The angle $\theta$ corresponds to the clock preparation (and is common to both wave packets) while the angle $\phi=\omega_0\Delta\tau$ describes the effect of the proper time lapse $\Delta\tau$ between the two clock wave packets, when the clock precesses at rate $\omega_0$ \cite{Zych,Us}. We consider the case where the distinguishability arises solely from $\Delta\tau$.  Because of imperfect clock preparation, $\Delta\tau$ may not increase the distinguishability $D$ to 1 (and correspondingly would not reduce the visibility to zero), and it is useful to characterize the actual distinguishability allowed by the clock by comparing it to the distinguishability $D_I$ made possible by a clock with an ideal preparation, where full distinguishability $D=1$ is achieved for $\Delta\phi\equiv \phi_u-\phi_d=\pi$, where $u$ and $d$ denote the upper and lower paths of the interferometer, respectively. We do this by introducing a re-scaling factor $C$ that accounts for such imperfection, taking $D=C\cdot D_I$.

\begin{figure}
\begin{center}
\includegraphics[width=16cm,height=11cm]{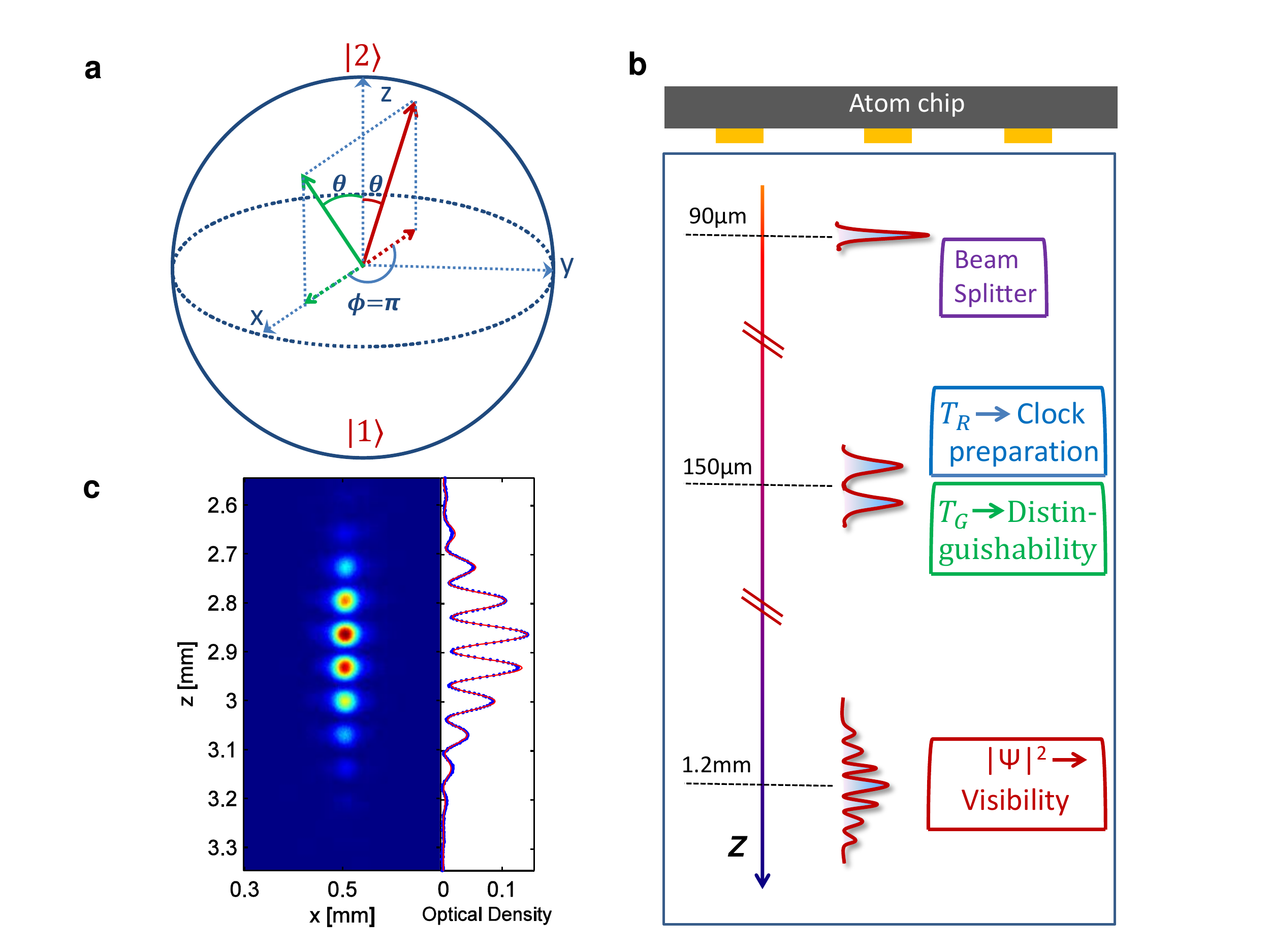}
\end{center}
\caption{ {\bf (a)} Bloch sphere of the clock interferometer, where the red (green) vector indicates the clock wave packet in the upper (lower) interferometer path.
The angle $2\theta$ between the two Bloch vectors (solid lines) is smaller than the angle $\pi$ between two vectors in a similar interferometer with a perfectly prepared clock (dashed lines).
{\bf (b)} Detailed experimental sequence (not to scale). $C(\theta$) is controlled by an RF pulse of duration $T_R$. $D_I(\phi)$ is controlled by a magnetic gradient pulse of length $T_G$.
{\bf (c)} 339 experimental shots of the interference pattern in a combined plot (one on top of the other, no alignment or corrections) when $D_I(T_G)=0$. The visibility is 0.789$\pm$0.001.  The mean of the single-shot visibility is 0.879$\pm$0.002. The errors are standard error of the mean (SEM).
}
\label{ExpScheme}
\end{figure}

Let us consider a clock that is initially prepared as a superposition $|\theta,\phi\rangle\equiv \cos(\theta/2)|1\rangle+e^{i\phi}\sin(\theta/2)|2\rangle$ of the two clock energy eigenstates $|1\rangle$ and $|2\rangle$. This clock state corresponds to a Bloch vector ${\bf s}=(\sin\theta\cos\phi, \sin\theta\sin\phi,\cos\theta)$, which is ideally at $\theta=\pi/2$ on the equator of the Bloch sphere, representing an equal superposition of the two energy eigenstates. After propagation along the two paths, the two clock wave packets acquire an angular difference $\Delta\phi=\omega_0\Delta\tau$ due to the proper time lag. The visibility $V$ of the clock interferometer is equal to the overlap $|\langle u|d\rangle|\equiv |\langle \theta,\phi_u|\theta,\phi_d\rangle|$ between the two states $|u\rangle$ and $|d\rangle$ of the clock wave packets, which have rotated angles $\phi_u$ and $\phi_d$, respectively, during free propagation at different heights in the gravitational field. The angular difference between the two states $|u\rangle$ and $|d\rangle$ makes them distinguishable; the interference visibility is reduced to zero if the overlap between the two states is zero, and the distinguishability $D\equiv \sqrt{1-|\langle u|d\rangle|^2}$ grows to 1, implying full ``which-path" information.
In the case of an ideal preparation, where $\cos(\theta/2)=\sin(\theta/2)=1/\sqrt{2}$, the angular separation between the two Bloch vectors ${\bf s}^u$ and ${\bf s}^d$ is $\Delta\phi=\phi_u-\phi_d$ and the overlap is $|\langle u|d\rangle| =|\cos(\Delta\phi/2)|$.
In general, we can choose two vectors ${\bf s}^a$ and ${\bf s}^b$ on the Bloch sphere, corresponding to two quantum states $|a\rangle$ and $|b\rangle$, with an angle of separation $\alpha_{ab}$ between them in the plane that they define.  Their overlap is likewise $\cos(\alpha_{ab}/2)$.
It follows that the distinguishability is
\begin{equation} D^2\equiv 1-|\langle a|b\rangle|^2=\sin^2(\alpha_{ab}/2)=\frac{1}{2}(1-\cos\alpha_{ab})=\frac{1}{2}(1-{\bf s}^a\cdot{\bf s}^b).
\label{eq:D2} \end{equation}
In our case, where the latitude $\theta$ of the clock states does not change over time, the (real) scalar product of the two Bloch vectors ${\bf s}^u$ and ${\bf s}^d$ is
${\bf s}^u\cdot{\bf s}^d=\sin^2\theta\cos\Delta\phi+\cos^2\theta$. We use the trigonometric equality $\cos\Delta\phi=1-2\sin^2(\Delta\phi/2)$ and note that $D_I =|\sin (\Delta \phi/2)|$ is the distinguishability of two states in an ideal clock prepared with the Bloch vector pointing to the equator, namely with equal populations.  Upon substituting ${\bf s}^u\cdot{\bf s}^d$ for ${\bf s}^a\cdot{\bf s}^b$ in Eq.~(\ref{eq:D2}) we obtain
\begin{equation} D^2=\sin^2\theta D_I^2,
\label{eq:D2theta} \end{equation}
namely, the distinguishability of the states of the two clock wave packets is a product of the distinguishability of two states created by perfect preparation of the clock and propagation through the interferometer, scaled by a factor $C=\sin\theta$, which varies from $C=1$ for an ideal clock to $C=0$ for a non-clock prepared in a given energy eigenstate (at the north or south pole of the Bloch sphere). While perfect clock preparation ($C=1$) gives rise to the possibility of perfect distinguishability $D=1$ (full orthogonality of the clock states) for a given proper time lag $\Delta\tau=\pi/\omega_0$, in the case of imperfect preparation ($C<1$) the angle between the Bloch vectors of the two wave packets is always smaller than $\alpha_{ud}=\pi$ and the maximum possible distinguishability is $D_{\rm max}=C<1$.
($C$ may be thought of as the clock preparation quality or ``clockness".)
In the context of a clock interferometer~\cite{Zych,Us}, where the distinguishability of clock states determines the visibility,
the complementarity relation $V^2+D^2\leq 1$ can now be written as
\begin{align}\begin{split}
V^2+(C \cdot D_I)^2 \leq  1~.
\label{eq:VCD}   \end{split}\end{align}
This is the clock complementarity relation, where $D_I$ is the ideal clock distinguishability, determined solely by the proper time lag $\Delta\tau$ (in the thought experiment).

The complementarity relation in Eq.~(\ref{eq:VCD}) was derived here for a typical atomic clock based on a two-level system. In this case the ideal distinguishability is $D_I(\Delta\tau)=|\sin(\omega_0\Delta\tau/2)|$ and the clock preparation quality is $C=\sin\theta=2\sqrt{P(1-P)}$, where $P$ and $1-P$ are the populations (occupation probabilities) of the two energy eigenstates of the clock.
In the more general case --- for example, a clock based on an $N$-level system [spin $S=(N-1)/2$] --- we show in Sect.~\ref{sec:multilevel} that Eq.~(\ref{eq:VCD}) leads to interesting results in which $\theta$ and $\phi$ may not be disentangled when defining $C$.

In the next section we demonstrate experimentally the complementarity relation of Eq.~(\ref{eq:VCD}) with a system of two Zeeman levels of an atom in a magnetic field. A vertical magnetic field gradient $\partial B/\partial z$ takes the place of the gravitational field. The accumulated angular difference between the two clock wave packets centered at heights $z_u$ and $z_d$ is $\phi_u-\phi_d=\Delta\omega_{\rm Zeeman} T_G$, where $T_G$ is the gradient pulse duration and $\Delta\omega_{\rm Zeeman}=g_F\mu_B(\partial B/\partial z)(z_u-z_d) /\hbar$, $\mu_B$ is the Bohr magneton and $g_F$ the Land\'e factor of the hyperfine level $F$. This clock shift mimics a shift $\omega_0\Delta\tau$ for a clock in two positions in the gravitational field, where $\Delta\tau\approx g(z_u-z_d)T/c^2$ and $T$ is the time (in the lab frame) during which the two wave packet centers are separated along the axis of gravity.

\section{Experimental scheme}
We experimentally verify to a high level of likelihood that a two-level clock obeys the generalized clock complementarity rule, {with a magnetic gradient simulating a gravitational red shift.} The setup used for this study is described in Ref. \cite{Us}, while numerous improvements have resulted in a much higher $V$ in the raw data: Fig. 1(c) shows very high visibility without any normalization (90\% as compared with 60\% in Ref. \cite{Us}). The experimental scheme is depicted in Fig. 1(b); it applies the previously demonstrated Stern-Gerlach (SG) matter-wave interferometer on an atom chip \cite{FGBS} in the following experimental sequence.  (For more details, see the Supplementary Material.) After a BEC of about $10^4$ $^{87}$Rb atoms in the state $\vert F, m_F\rangle =\vert 2,2\rangle$ has been released from a magnetic trap located 90$\pm2$ ${\mu}$m below the chip surface, the SG beam splitter acts on it.
\begin{figure}
\begin{center}
\includegraphics[width=16cm,height=11cm]{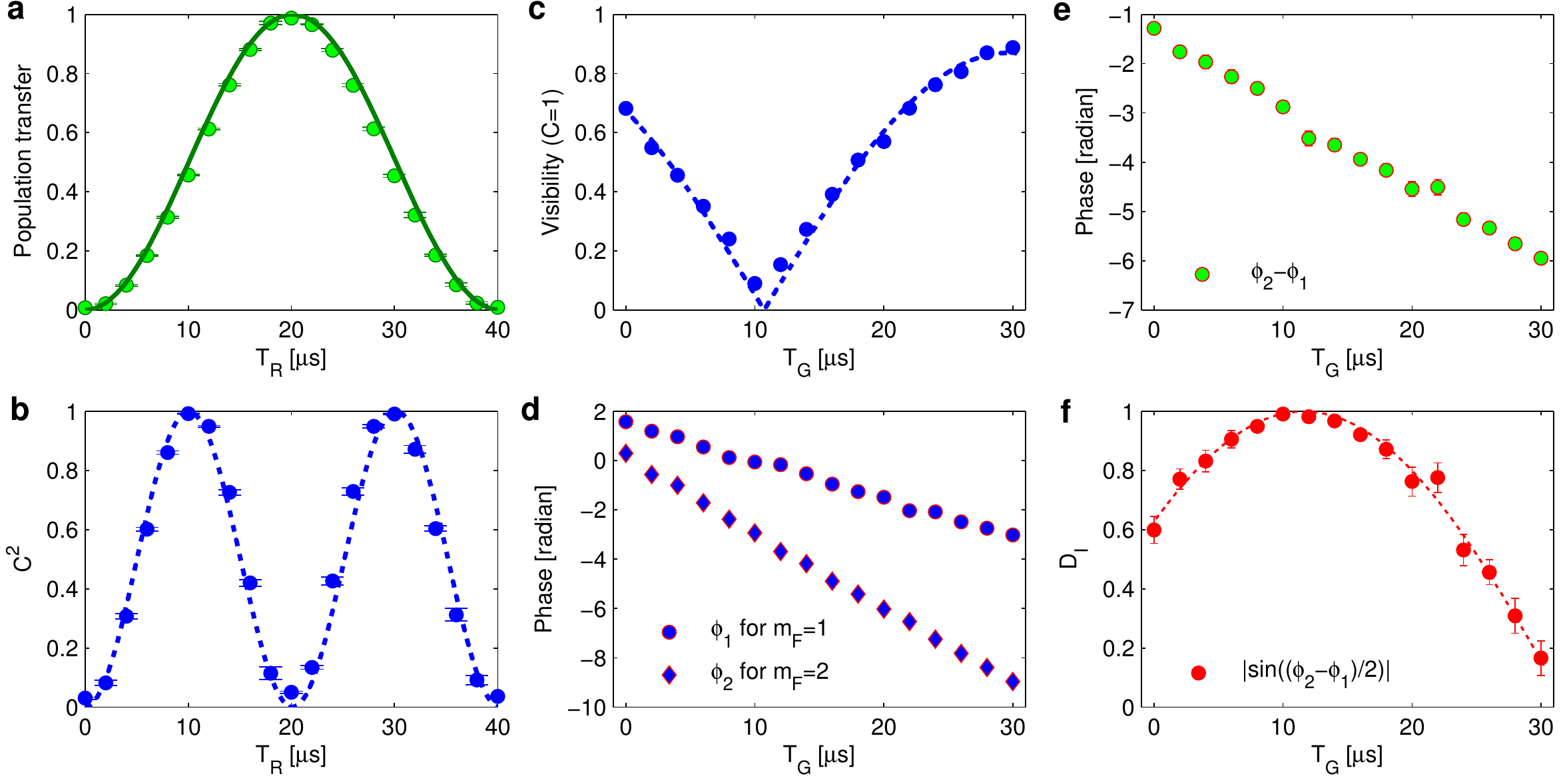}
\end{center}
\caption{ The independent measurement of $V$, $C$ and $D_I$: {\bf (a,b)} $C^2 = 4P(1-P)$ is measured independently in a separate experiment by measuring the population transfer $P$ after the clock is prepared by an RF pulse of duration $T_R$; {\bf (c)} the visibility of an ideal clock ($C=1$) interference pattern versus $T_G$, which induces distinguishability; the result is fitted to $|\cos(\phi/2)|$. {\bf (d-f)} $D_I$ is evaluated independently by measuring the relative angle in two single-state interferometers each containing one of the two clock states, $\phi_1$ for $m_F=1$ and $\phi_2$ for $m_F=2$, and then by calculating $D_I=|\sin(\phi_2-\phi_1)/2|$. The errors are standard error of the mean (SEM) and are at times not visible because of their small magnitude.
}
\label{Experiment}
\end{figure}
It creates a coherent spatial superposition of two wave packets in the same spin state ($\vert 2,2\rangle$). A stopping pulse then adjusts the relative velocity of the two wave packets so that they have the same momentum. Clocks are prepared by an RF pulse of duration $T_R$, which creates a superposition of $\vert 2,2\rangle\equiv\vert 2\rangle$ and $\vert 2,1\rangle\equiv\vert 1\rangle$ states. The pulses are applied under a strong homogeneous magnetic field (36.7 G) in order to push the transition to $\vert2,0\rangle$ out of resonance via the nonlinear Zeeman effect, thus forming a pure two-level system. As the Rabi frequency $\Omega_{R}$ is constant, varying $T_R$ will effectively change the Bloch vector's polar angle $\theta$ in the Bloch sphere [Fig. 1(a)], e.g. when $T_R=0\,\mu$s, there is no rotation and the Bloch vector stays at the north pole, and when $T_R=10\,\mu$s, the Bloch vector is rotated onto the equator and a proper clock is prepared in the state $(\vert 2\rangle+\vert 1\rangle)/\sqrt{2}$. Then an additional magnetic gradient pulse of duration $T_G$ is applied in order to change the relative ``tick" rate of the superposed clock wave packets, thus determining a relative rotation $\phi$ on the equator of Bloch sphere [Fig. 1(a)]. This synthetic red shift introduces {\it a posteriori} which-path information (WPI) by creating entanglement between the path and a WPI marker, in contrast to the {\it a priori} WPI, which involves the preparation of an unbalanced interferometer such that the particle flux along the two paths differs.

Let us note that it is not enough to experimentally simulate the thought experiment by placing a clock in a spatial superposition, and creating a synthetic red shift with some force field. To faithfully simulate the thought experiment one must make sure that there is no breakup of the clock due to the applied force field. This may be viewed as a mere technical condition for the operation of a clock, but in fact the ``no clock breakup" is a fundamental feature of the thought experiment that must be imitated by any experimental simulation. Specifically, there is no breakup of a clock wave packet in the gravitational field.  Consider a {\it single} wave packet centered at a point $z_0$ and let $\tau (z_0)$ be a proper time lapse there.  While two clock levels are indeed accelerated in the gravitational field to different momenta $p_j=m_j g\tau(z_0)$ (where the mass difference $m_2 - m_1 = \hbar\omega_0/c^2$ is due to the difference $\hbar\omega_0$ in their energies), the corresponding velocities $v_j=p_j/m_j=g\tau(z_0)$ do not depend on the specific clock level.  The Galilean law of falling masses, stating that gravitational acceleration is independent of mass, holds in general relativity and insures that clock breakup will not occur in a gravitational field.
Similarly, the clock breakup effect in our experiment is negligible relative to the difference of the clock angle between the two clock wave packets.
A clock wave packet $\psi_0(z)(|1\rangle+|2\rangle)$ in a magnetic field gradient undergoes not only a rotation of the clock $|1\rangle+|2\rangle\to |1\rangle e^{-i\omega_1 T_G}+|2\rangle e^{-i\omega_2 T_G}$, where $\omega_1$ and $\omega_2$ are the magnetic potentials for the two levels at $z_0$, but also a differential momentum.
While the differential clock rotation is shown to span a large range of clock angles allowing the two clock states to be fully distinguishable ($D_I=1$), the momentum separation between the two states of the same clock, which is given by $\Delta p=\hbar(\partial \omega_1/\partial z-\partial\omega_2/\partial z)T_G$, is much smaller than the momentum distribution of each wave packet (allowing the observation of many spatial fringes \cite{Us}). These conditions are automatically fulfilled in our experiment when the separation between the two wave packets is larger than the wave packet width. It follows that our demonstration of the effect of gravitational red shift on clock distinguishability is valid.

\section{Verifying clock complementarity}
Each clock is a superposition of two Zeeman sublevels, with coefficients that depend on $\theta$ and $\phi$.  The RF pulse (duration $T_R$) controls the value of $C=\sin\theta$, while the magnetic gradient pulse (duration $T_G$) controls the value of $D_I=\sin(\phi/2)$. The latter creates an effective red shift, namely a differential clock ``tick" rate, by inducing a differential Zeeman splitting $\Delta\omega$ such that $\phi=\Delta\omega\cdot T_G$. Finally $V$ is measured from the spatial interference pattern [Fig. 1(c)]. We measure $C^2=4P(1-P)$ independently in a separate experiment by measuring $P$ after the clock is initialized, and we evaluate $D_I$ independently by measuring the relative phases in two single-state interferometers, one for each of the two clock states. The independent measurements of $V$, $C$ and $D_I$ are presented in Fig. 2.

As noted, $C = \sin \theta = 2\sqrt{P(1-P)}$, and in order to establish the value of $C$ we need to measure the population transfer from the $m_F = 2$ state to the $m_F = 1$ state.
In Fig. 2(a)
we show the population transfer measured by Stern-Gerlach splitting of the different spin states and atom counting, and Fig. 2(b)
shows the resulting value of $C^2$. As expected, $C^2$ oscillates between 0 and 1, corresponding to the population transfer.
In Fig. 2(c),
we scan $T_G$ and measure the optimal clock ($C=1$) interference visibility. The result is fitted with $\vert \cos(\phi/2)\vert$, where $\phi$ represents the clock relative rotation.
In Figs. 2(d-f)
$D_I$ is measured by two single-state interferometers ($m_F=2$ and $m_F=1$). The difference in phase between these two interferometric fringe patterns is equivalent to the relative rotation $\phi$ between the upper and lower clock wave packets, from which $D_I$ is directly calculated as $\vert \sin(\phi/2)\vert$.  The measured relations among the population transfer $P$, the parameter $C^2$ and the visibility $V$ appear in greater detail in Fig. 3, for the case of $D_I$ equal to 1.

\begin{figure}
\begin{center}
\includegraphics[width=16cm, height=14cm]{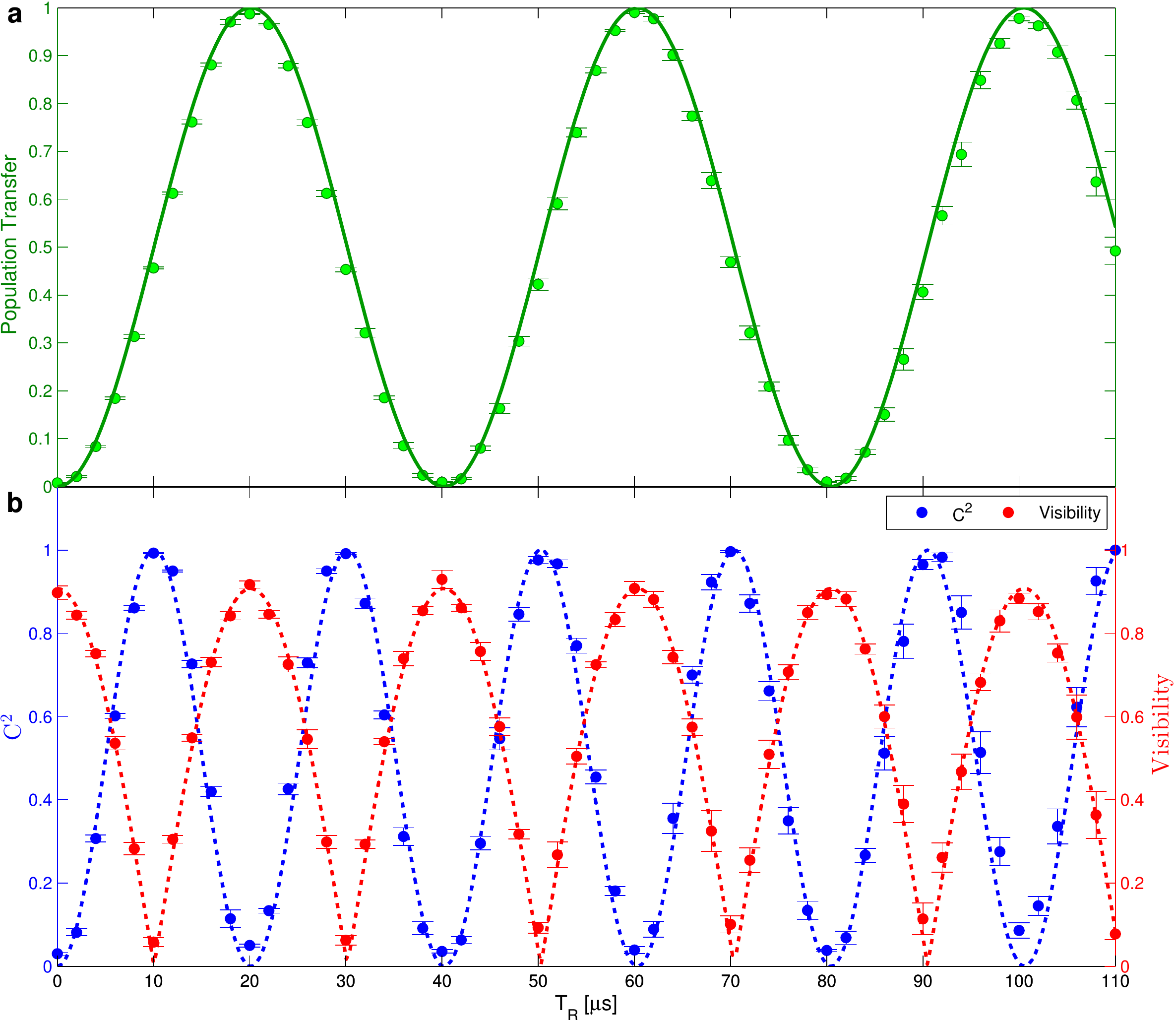}
\end{center}
\caption{{\bf (a)} Clock preparation, showing the population transfer $P$ vs. $T_R$. {\bf (b)} In blue, the measured $C^2$  vs. $T_R$, when $\Delta \phi=\pi$ (and $D_I=1$ with an uncertainty of 1\%), as well as (dashed line) the calculated $C^2=4P(1-P)$, taking $P$ from {\bf (a)}.  For reference, we also show (in red) the measured $V$ vs. $T_R$, as well as (dashed line),  the calculated visibility $V=|\cos\theta|V_{max}=|1-2P| V_{max}$ [again, taking $P$ from {\bf (a)}], where $V_{max}=0.9$ is our maximal visibility limited by optical resolution, etc.
The figure shows the complementary between $C^2$ and $V$ when $D_I$ equals 1.
}
\label{fidvis2}
\end{figure}

In Fig. 4(a) we present the clock complementarity relation $V^2+(C \cdot D_I)^2$ for four values of $C$ when $D_I$ is scanned. Fig. 4(b) presents the clock complementarity for four values of $D_I$ when $C$ is scanned. With $V$, $C$ and $D_I$ measured independently, Fig. 4 demonstrates that the clock complementarity rule is sound.


\begin{figure}
\begin{center}
\includegraphics[width=12.cm, height=10.cm]{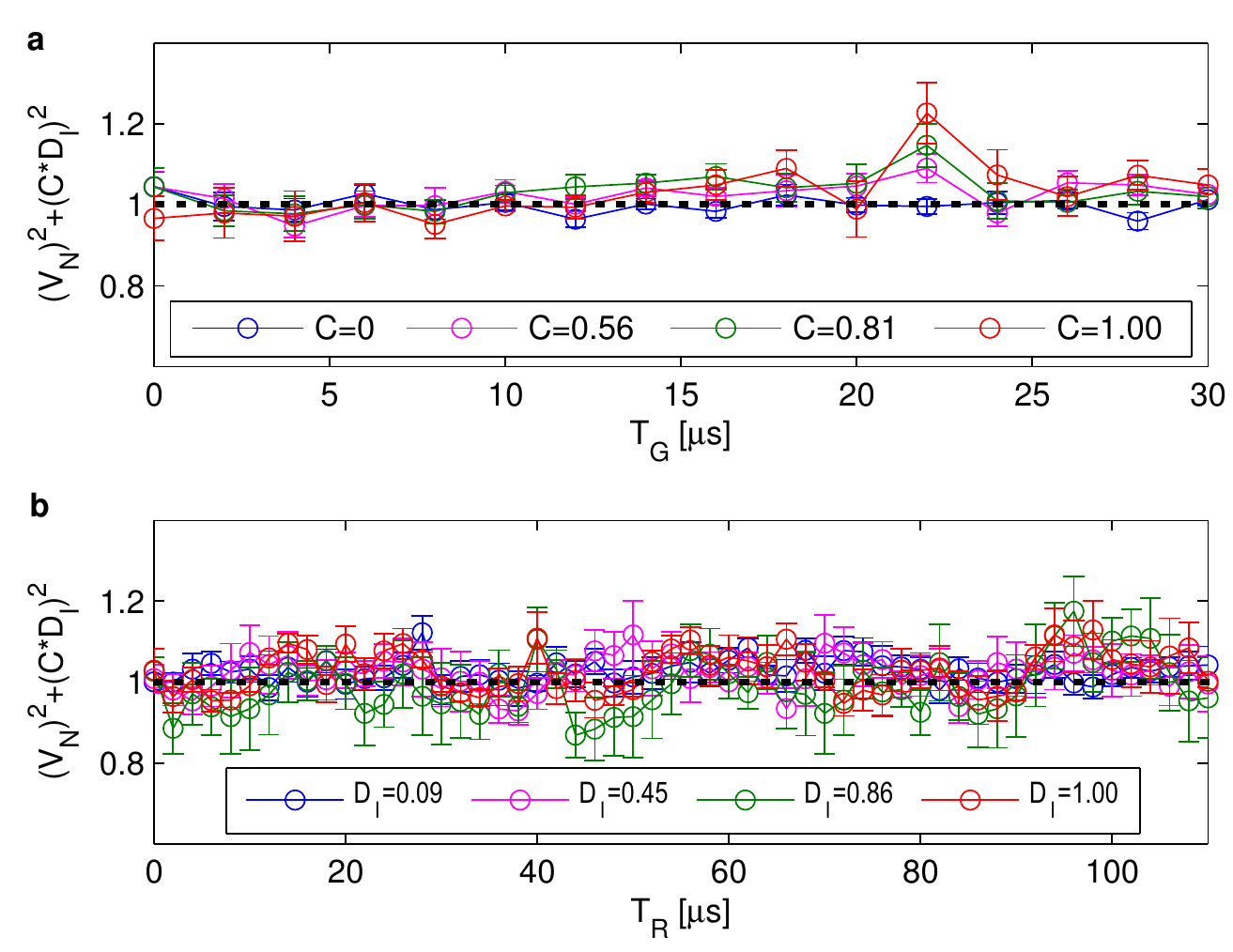}
\end{center}
\caption{The value of $V^2+(C\cdot D_I)^2$, where all three parameters are measured independently: {\bf (a)} for four values of $C$ when $D_I$ is scanned, and
{\bf (b)} for four values of $D_I$  when $C$ is scanned. $V$ is the normalized visibility:
Each value of the visibility is an average of the single-shot visibility from several experimental cycles, and the error bars represent the standard error of the mean (SEM) in this sub-sample. For error bars corresponding to standard deviation (SD) we multiply by $\sqrt{n}$, where $n=6$ is the number of data points.  This average is normalized to the visibility of the single-state interferometer (i.e. without an initialization of a clock) to account for experimental imperfections.
}
\label{fidvis1}
\end{figure}

\begin{figure}
\begin{center}
\includegraphics[width=17.5cm, height=10cm]{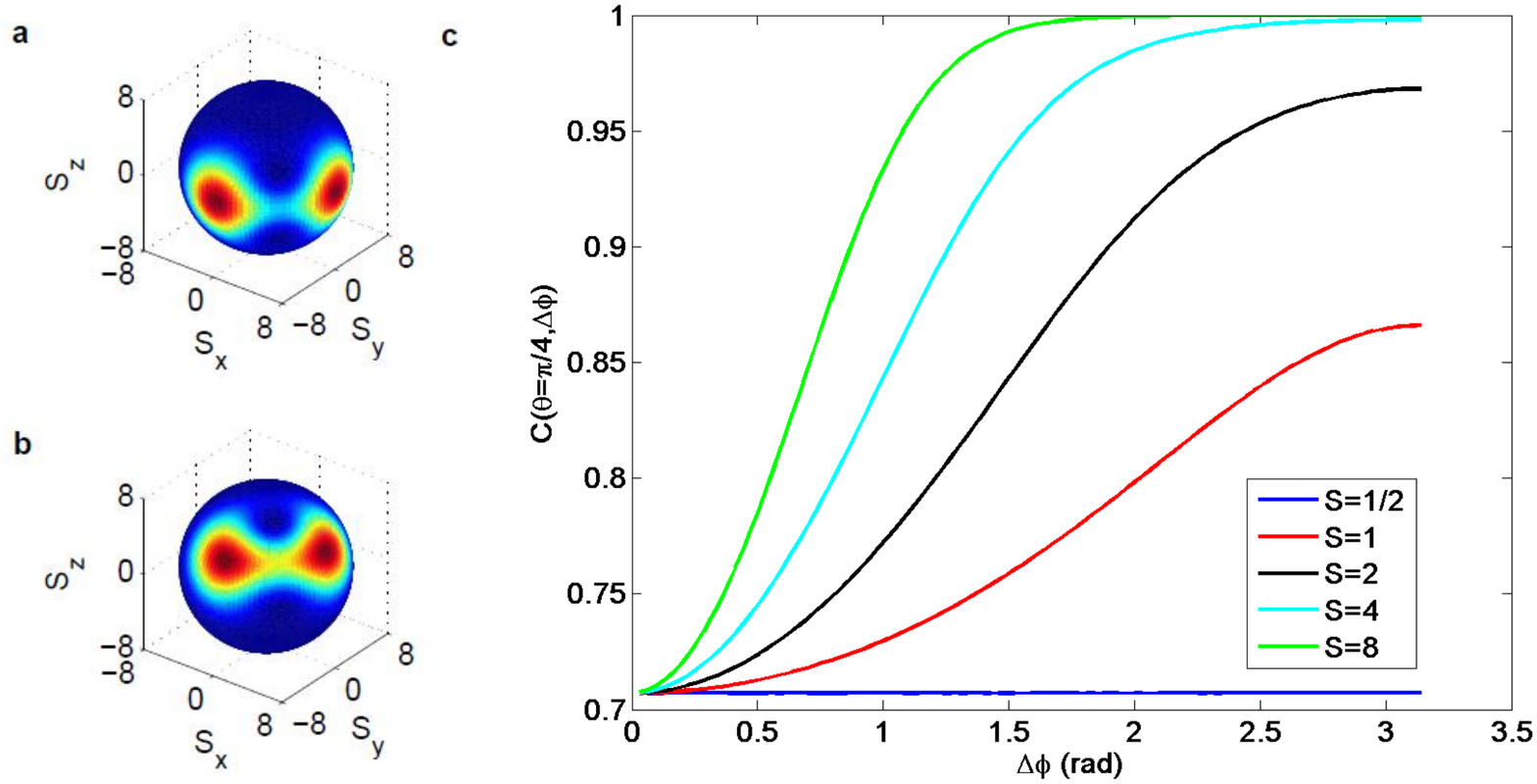}
\end{center}
\caption{
Distinguishability for coherent states of spin $S\geq 1/2$. {\bf (a)} The Bloch sphere of $S=8$ showing the angular distribution of a superposition of states $|\theta,\phi\rangle=|\pi/2,0\rangle$ and $|\pi/2,-\pi/2\rangle$ with almost full distinguishability. {\bf (b)} A similar superposition for a non-ideal clock prepared at $\theta=\pi/3$; the two states show a considerable overlap.
{\bf (c)} ``Clockness" $C$ for a preparation angle $\theta=\pi/4$ as a function of the phase difference $\Delta\phi$ and different spin values $S$. For $\Delta\phi\to 0$ $C\to \sin\theta$ is independent of spin, but for large $\Delta\phi$ the ``clockness" is large for large spins as the angular distribution on the Bloch sphere is narrow, implying high distinguishability regardless of the preparation angle.}
\label{fig:multilevel}
\label{fidvis3}
\end{figure}

\section{Multilevel clocks}
\label{sec:multilevel}

To achieve an atomic clock with a better time precision it is possible to choose a pair of energy eigenstates with a larger energy spacing $\hbar\omega_0$. In the context of our Zeeman level clock it is possible, for example, to prepare the system as a superposition of the two extreme Zeeman levels $m_F=\pm 2$ of the $F=2$ manifold and use this system as a two-level clock with rotation frequency $2F\omega_0$. (See \cite{Lovecchio2016} for a possible realization.) The discussion in Sect.~\ref{sec:theory} is valid for this system exactly in the same way.

An example of a multilevel clock where a few or many levels are occupied simultaneously during the clock evolution provides a model for examining the transition to the classical clock limit where the clock hand moves over a continuum of distinguishable times.
So far, a two-level clock was prepared by using a Rabi rotation that places the $S=1/2$ Bloch vector at an angle $\theta$ from the $z$ axis of the Bloch sphere. Let us consider an $S> 1/2$ system prepared in a similar way. (For an example of such a preparation see~\cite{Petrovic2013}; for a possible realization of a very large $S$ see \cite{Budker}.)
Figs.~\ref{fig:multilevel}(a,b) show an $S=8$ clock interferometer on the Bloch sphere. In a spin-$S$ system (with $N=2S+1$ levels and equal energy spacing), one may rotate the state along the $\theta$ direction while free evolution rotates the state along the $\phi$ direction. As in the spin-$1/2$ system, the overlap between two states $|\theta_a,\phi_a\rangle$ and $|\theta_b,\phi_b\rangle$, representing two coherent states obtained by such rotations starting from the extreme energy eigenstate $m_S =S$, is determined by the angle $\alpha_{ab}$ between the two Bloch vectors ${\bf s}^a$ and ${\bf s}^b$ corresponding to the two quantum states. For example, consider the overlap between the two states $|\theta_a,\phi_a\rangle=|0,0\rangle$ (the extreme energy eigenstate on the north pole) and $|\theta,\phi\rangle$ obtained by a Rabi rotation of the state $|0,0\rangle$ with an angle $\theta$. This state has the form
\begin{equation} |\theta,\phi\rangle=\sum_{m=-S}^S \cos^{S+m}(\theta/2)\sin^{S-m}(\theta/2)\sqrt{\left(\begin{array}{c} 2S \\ S+m\end{array}\right)}e^{-im\phi}|S,m\rangle, \end{equation}
where $|S,m\rangle$ are the spin eigenstates and $\left(\begin{array}{c} 2S \\ S+m\end{array}\right)$ are binomial coefficients for choosing $S+m$ out of $2S+1$. It follows that the overlap integral is given by $|\langle 0,0|\theta,\phi\rangle|=\cos^{2S}(\theta/2)$. As rotations around the Bloch sphere are unitary operations and do not change the overlap between two states transformed under the same operation, and as can be verified directly from the above equation, we can generalize this result to any two coherent states on the Bloch sphere, such that
\begin{equation} |\langle \theta_a,\phi_a|\theta_b,\phi_b\rangle|=\cos^{2S}(\alpha_{ab}/2),
\end{equation}
where $\alpha_{ab}$ is the angle between the two Bloch vectors such that $\cos\alpha_{ab}={\bf s}^a\cdot{\bf s}^b$. By using some trigonometric equations we conclude that for two Bloch vectors prepared at the same latitude $\theta$ the distinguishability is
\begin{equation} D^2=1-\left[\frac{1}{2}(1+{\bf s}^u\cdot{\bf s}^d)\right]^{2S}
=1-[1-\sin^2\theta\sin^2(\Delta\phi/2)]^{2S}. \end{equation}
For $S=1/2$ this leads to the same expression as in Eqs.~(\ref{eq:D2}) and~(\ref{eq:D2theta}). The ideal distinguishability is $D_I^2=1-\cos^{4S}(\Delta\phi/2)$ (conforming to the two-level system result for $S=1/2$). This implies that the ``clockness" $C$ should be
\begin{equation} C^2\equiv \frac{D^2}{D_I^2}=
\frac{1-[1-\sin^2\theta\sin^2(\Delta\phi/2)]^{2S}}{1-\cos^{4S}(\Delta\phi/2)}.
\end{equation}
In the limit of a very short time lag $\Delta\phi\to 0$, the ``clockness" becomes $C\to \sin\theta$, the same as for spin-$1/2$ and independent of the spin. However, for general proper time lags of the two clocks, $C$ becomes dependent both on the spin $S$ and the angle difference $\Delta\phi$. The value of $C$ as a function of $\Delta\phi$ is shown in Fig.~\ref{fig:multilevel}(c).
For large values of the spin $S$ and large proper time differences, the distinguishability is no longer sensitive to the clock preparation angle, as the clock states are represented by a narrow distribution of angles on the Bloch sphere and therefore two states with large $\Delta\phi$ are well separated even if the preparation angle is not ideal.


Finally, we can apply the clock complementarity relation in Eq.~(\ref{eq:VCD}) to a single-state spatial interferometer, e.g. the Compton clock for which $C=0$; \cite{Muller4,CCT2,Wolfgang2,Peil}; but $C=0$ does not correspond to a clock in the usual sense of an internal state space. What is unique to $C>0$ clock interferometry is the reduced $V$ due to different clock readings along the paths, rendering the paths distinguishable \cite{Zych,Us}. An additional implication of Eqs. (1-2) is that $V^2+D^2 > 1$ \cite{Zych} requires either $V\ne|\langle {\bf s}^u |{\bf s}^d\rangle|$ or new rules for scalar products in quantum mechanics.

\section{Conclusion}
In summary, we have theoretically obtained and experimentally confirmed a clock complementarity relation, $V^2+(C \cdot D_I)^2 = 1$, for clock wave packets superposed on two paths through an interferometer.  Here $V$ is the visibility of their interference pattern, $C$ is a measure of the ``preparation quality" of the clock, and  $D_I$ is the distinguishability of an ideally prepared clock. We emphasize that our experiment measures $V$, $C$, $D_I$ independently.  While this relation is specific to clock complementarity, it is unusual in linking non-relativistic quantum mechanics with general relativity.  A direct test of this complementarity relation will come when $D_I$ reflects the gravitational red shift between two paths which traverse different heights.

\newpage

{\bf Acknowledgments}

We gratefully acknowledge discussions with Wolfgang Schleich and Albert Roura.
We thank Zina Binstock for the electronics and the BGU nano-fabrication facility for providing the high-quality chip.
This work is funded in part by the Israel Science Foundation (grant no. 1381/13), the EC Matter--Wave consortium (FP7--ICT--601180), and the German-Israeli DIP
project (Hybrid devices: FO 703/2--1) supported by the DFG. We also acknowledge support from the Israeli Council for Higher Education and from the Ministry of Immigrant Absorption (Israel). D. R. thanks the John Templeton Foundation (Project ID 43297) and the Israel Science Foundation (grant no. 1190/13) for support. The opinions expressed in this publication do not necessarily reflect the views of the John
Templeton Foundation.

\bigskip
\newpage


\pagebreak


\renewcommand{\vec}[1]{\boldsymbol{\mathbf{#1}}}
\newcommand{\phimag}{\phi_{\text{mag}}}
\newcommand{\mm}{\text{mm}}
\renewcommand{\theequation}{S\arabic{equation}}
\renewcommand{\thefigure}{S\arabic{figure}}
\setcounter{figure}{0}

\centerline{Supplementary Materials for}
\bigskip
\bigskip

\centerline{\bf Quantum complementarity of clocks in the context of general relativity}

\centerline{Zhifan Zhou, Yair Margalit, Daniel Rohrlich,$^*$ Yonathan Japha and Ron Folman}

\author{Zhifan Zhou}	
	\affiliation{Department of Physics, Ben-Gurion University of the Negev, Be'er Sheva 84105, Israel}
\author{Yair Margalit}
	\affiliation{Department of Physics, Ben-Gurion University of the Negev, Be'er Sheva 84105, Israel}	
\author{Daniel Rohrlich}
	\affiliation{Department of Physics, Ben-Gurion University of the Negev, Be'er Sheva 84105, Israel}	
\author{Yonathan Japha}
	\affiliation{Department of Physics, Ben-Gurion University of the Negev, Be'er Sheva 84105, Israel}	
\author{Ron Folman}
	\affiliation{Department of Physics, Ben-Gurion University of the Negev, Be'er Sheva 84105, Israel}


\centerline{$^\ast$Corresponding author. E-mail: rohrlich@bgu.ac.il}
\bigskip
\bigskip
\bigskip
\bigskip
\bigskip
{\bf This PDF file includes:}

\begin{itemize}
  \item[] Experimental Methods (Fig.\,1)
  \item[] {\it a priori} and {\it a posteriori} which-path information
  \item[] Properties of visibility
  \item[] Verifying the clock complementarity rule (Fig.\,4)
  \item[] Measuring $C$ and $D_I$ independently (Fig.\,2a-b and Fig.\,2d-f)
  \item[] Applying the new rule to the Compton clock debate
  \item[] Breaking the complementarity bound
  \item[] No-clock interpretations
\end{itemize}

\pagebreak

{\bf S1. Experimental Methods (Fig.\,1)}

We start our experiment by preparing two wave packets in a spatial superposition via the previously demonstrated Stern-Gerlach (SG) type matter-wave interferometer on an atom chip \cite{FGBS}. The SG beam splitter (SGBS) is applied after a BEC of about $10^4$ $^{87}$Rb atoms in the state $\vert F, m_F\rangle =\vert 2,2\rangle$ is released from a magnetic trap located 90 $\pm2$ ${\mu}$m below the chip surface. The trap is created by a copper structure located behind the chip with the help of additional homogeneous magnetic bias fields in the x, y and z directions (See Fig. S1). The SGBS includes a first radio-frequency (RF) $\pi/2$ pulse (of 10 $\mu$s duration) to create an internal spin-state superposition composed of  $|1\rangle\equiv \vert 2,1\rangle$ and $|2\rangle\equiv \vert 2,2\rangle$, a magnetic gradient pulse (4 $\mu$s) which creates a different magnetic potential for different spin states, and another RF pulse (10 $\mu$s).  These pulses create a superposition of $|2\rangle$ wave packets having different momenta (as well as a superposition of $|1\rangle$ wave packets, which we choose to discard). After the SGBS, we apply a second magnetic gradient of 90-98 $\mu$s duration to zero the relative velocity of the wave packets.

Following the above splitting procedure, clocks are initialized in both wave packets 1.5 ms after trap release by a third RF pulse of duration $T_R$, which creates a superposition of the $\vert 2\rangle$ and $\vert 1\rangle$ states. As the Rabi frequency $\Omega_{R}$ is constant, varying $T_R$ will effectively change the Bloch vector's rotation $\theta$  (Fig.\,1a), e.g. when $T_R=0~\mu$s, there is no rotation and the Bloch vector stays at the north pole, and when $T_R=10~\mu$s, the Bloch vector is rotated onto the equator and a proper clock is prepared in the state $(\vert 2\rangle+\vert 1\rangle)/\sqrt{2}$. Then an additional (third) magnetic gradient pulse of duration $T_G$ is applied in order to change the relative ``tick" rate of the two clock wave packets, corresponding to a relative rotation angle $\phi$ along the equator of the Bloch sphere (Fig.\,1a), simulating the effect of the red shift.

The entire SGBS and clock initialization sequences are done under a strong homogeneous magnetic field of 36.7 G in the $\hat y$ direction, which creates an effective two-level system via the non-linear Zeeman effect ($E_{21}$ $\approx h\times$25\,MHz, $E_{21}-E_{10}  \approx h\times $180\,kHz, where $E_{ij}$ is the energy difference between level $i$ and $j$). This field is adiabatically turned off 3.5 ms after the clock initialization (5 ms after trap release), leaving earth's magnetic field to preserve the two-level system continuously. After an additional 11-13 ms time-of-flight (16-18 ms after trap release, where the different times correspond to different optimization depending on the parameter we scan), we image the atoms by absorption imaging and generate the picture shown in Fig.\,1c.

\begin{figure}
\centerline{
\includegraphics*[width=\textwidth]{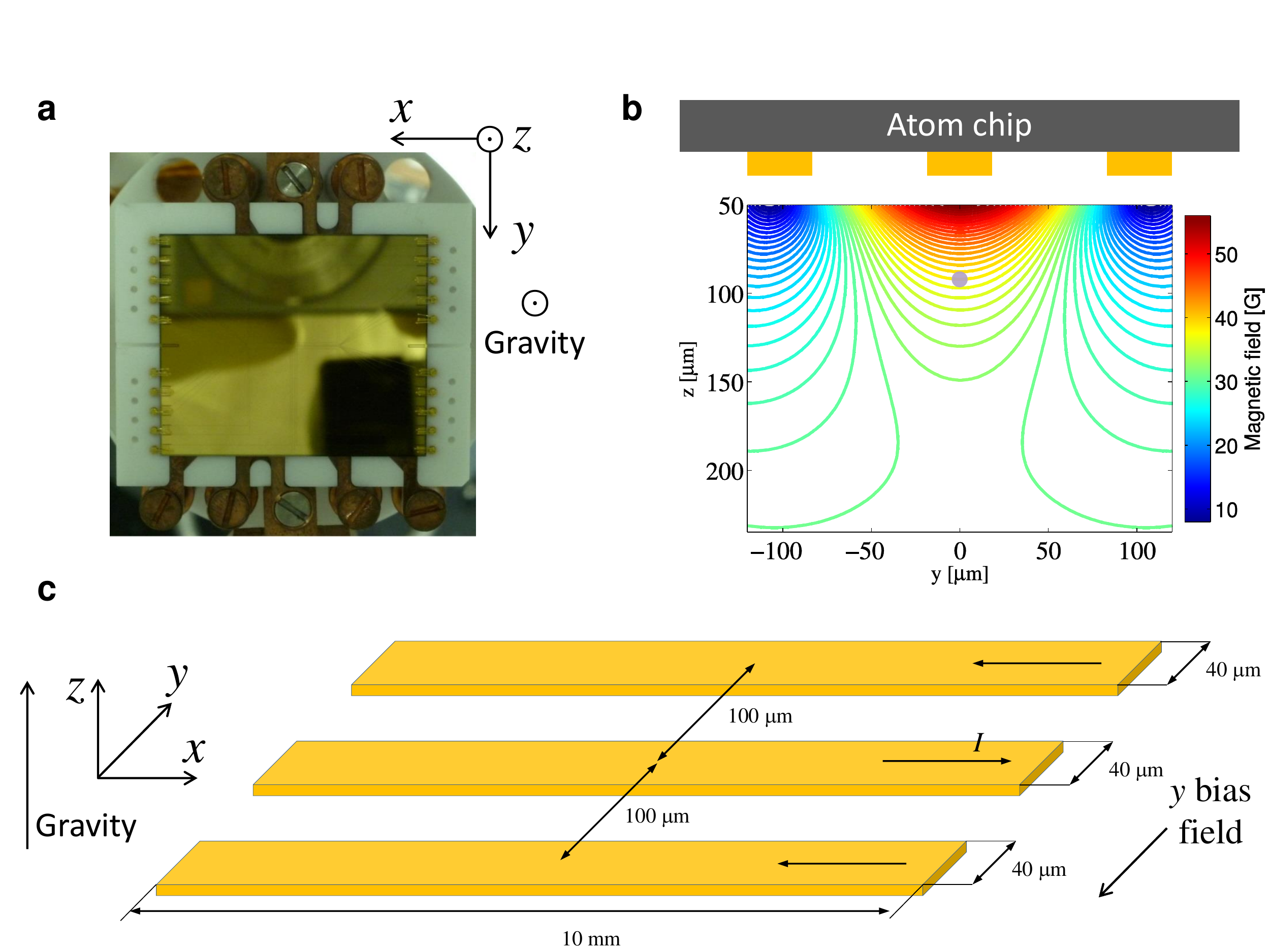}}
\caption{{\bf a}, A picture of the atom chip on its mount, with the copper structure visible behind it. Note that its orientation in the experimental setup is face down. {\bf b}, Magnetic field strength below the atom chip, generated by the quadrupole field via the chip wires and the bias field $B_y$ via external coils. The purple dot shows the location of the trapped BEC, which has, according to simulation, a Thomas-Fermi half-width in the $yz$ plane of about 3 $\mu$m. {\bf c}, Schematic diagram of the relevant chip wires. Wires are 10 mm long, 40 $\mu$m wide and 2 $\mu$m thick. The separation of the wires' centres is 100 $\mu$m, and the direction of the current $I$ alternates from one wire to the next.  The wires, being much smaller than the chip, are hardly visible in {\bf a}.}
\end{figure}

All three magnetic gradient pulses are generated by three parallel gold wires located on the chip surface (Fig. S1), which are 10 mm long, 40 $\mu$m wide and 2 $\mu$m thick.  The wires' centres are separated by 100 $\mu$m, and the same current runs through them in alternating directions, creating a 2D quadrupole field at $z$ = 98 $\mu$m below the atom chip (including finite size effects). The SGBS phase noise is largely proportional to the magnitude of the magnetic field created during the gradient pulse \cite{FGBS}. As the main source of magnetic instability is in the gradient pulse originating from the chip, positioning the atoms near the middle (zero) of the quadrupole field created solely by the three chip wires 98 $\mu$m below the chip surface reduces the phase noise during the SGBS operation. The chip wire current was driven using a simple 12.5 V battery, and was modulated using a home-made current shutter, with ON/OFF times as short as 1 $\mu$s.  The total resistance of the three chip wires is 13.6 $\Omega$, yielding a current of 11.3/13.6 A $\approx 0.83$ A. (A small voltage drop exists in the circuit itself.)

The RF signal is generated by an Agilent 33250A waveform generator and subsequently amplified by a Minicircuit ZHL-3A amplifier.  We generate RF pulses using a Minicircuit ZYSWA-2-50DR RF switch.  RF radiation is transmitted through two of the copper wires located behind the chip (with their leads showing in Fig. S1).

Fig.\,1c includes 339 experimental shots in a combined plot (no alignment or corrections) when $D_I(T_G)=0$. The visibility is 0.789$\pm$0.001.  The mean of the single-shot visibility is 0.879$\pm$0.002.
To verify that for $D_I(T_G)=0$ the value of $C(T_R)$ has no influence, $T_R$ was varied from 0 to 110\,$\mu$s in time steps of 2\,$\mu$s. As 6 shots were taken in each time step, the total number of images comes out to be $(110/2+1)\times 6=336$ where an additional 3 were taken at $T_R=0$.

\bigskip

{\bf S2. {\it a priori} and {\it a posteriori} which path information}

Quantum complementarity is basic to our understanding of QM \cite{Bohr}. The complementarity principle is typically studied by means of interferometers, e.g. a Mach-Zehnder interferometer, to examine the mutual exclusiveness of distinguishability $D$ (also known as ``which path'' information, WPI, or ``particle property'' $P$) between the different wave packets, and interferometric visibility $V$ (also known as ``wave property'', $W$). Following the work of Greenberger, Yasin, Englert, Jaeger, Shimony, Vaidman and others, our present understanding may be summarized by the fundamental inequality $V^2+D^2 \leq 1$  \cite{Englert, Greenberg, Jaeger, Mandel, Zurek}. This inequality is at times defined as $W^2+P^2\leq1$, in which case it is said to describe wave-particle duality. This law of complementarity has been verified in numerous experiments \cite{Rauch, Aspect, Rempe, MandelExp, Haroche2, Electron, Pfau, Chapman}.

Let us briefly note that it is commonly suggested to test complementarity by two different methods: the first requires the preparation of an unbalanced interferometer such that the particle flux along the two paths differs, and this creates {\it a priori} which-path information\,(WPI) as we have some information regarding which path a particle took\,(also called predictability); the second utilizes a balanced interferometer but introduces {\it a posteriori} WPI by creating entanglement between the path and a WPI marker or by using an interferometer with an unbalanced output beam splitter \cite{Aspect,Rempe}. In our thought experiment as well as in the experimental demonstration, a balanced interferometer is used where the {\it a posteriori} WPI is created by proper time, which we simulate with a synthetic gravitational red shift in the form of a magnetic gradient.

\bigskip
{\bf S3. Properties of visibility}

In this section we prove two fundamental features of visibility. The first is that it is equal to the overlap of the states in the two paths, and the second is that it is not dependent on the initial clock state, i.e. on the phases among the internal atomic spin states as the atom enters the interferometer.

The standard definition of the visibility $V$ of an interference pattern is
\begin{equation}
V= {{{\rm max} -{\rm min}}\over {{\rm max} +{\rm min}}}~~~,
\end{equation}
where ``max" and ``min" refer, respectively, to the densities in the peaks (maxima) and troughs (minima) of the interference pattern.  These densities are proportional to the absolute value squared of the atomic wave function, the superposition of the two normalized clock wave packets $\vert u\rangle$ and $\vert d\rangle$.  However, the relative phase of $\vert u\rangle$ and $\vert d\rangle$ in the superposition depends on their spatial position, and ``max" and ``min" refer to those positions in space where the densities are extremal.  A convenient way to take this phase into account is to write the superposition as $\vert u\rangle +e^{i\chi} \vert d\rangle$ and then find the value of $\chi$ that yields the visibility.  Note that there is only one phase $\chi$ to calculate, since if $\vert \chi_+ \rangle \equiv \vert u\rangle +e^{i\chi} \vert d\rangle$ corresponds to a peak, then $\vert \alpha_- \rangle \equiv\vert u\rangle -e^{i\chi} \vert d\rangle$ corresponds to a trough. (Note that $\vert \chi_+ \rangle$ and  $\vert \chi_- \rangle$  are not normalized.) We thus define $V_\chi$ as
\begin{eqnarray}
V_\chi &=& {{{\langle \chi_+\vert \chi_+\rangle} -{\langle \chi_-\vert \chi_-\rangle}}\over {{\langle \chi_+\vert \chi_+\rangle} +{\langle \chi_-\vert \chi_-\rangle}}}\cr
&&\cr
&=& {1\over 2} \left[ e^{i\chi}\langle d\vert u\rangle +
e^{-i\chi}\langle u\vert d\rangle\right] ~~~,
\end{eqnarray}
which equals the real part of $e^{i\chi}\langle d\vert u\rangle$.  The real part of $V_\chi$ is maximal when $e^{i\chi}\langle d\vert u\rangle$ equals $\vert\langle d\vert u\rangle\vert$, hence $V= \vert \langle d\vert u\rangle\vert$.

To prove that $V$ does not depend on the initial clock state, we restrict ourselves to pure states and assume that the magnetic field along each path is homogeneous, such that the spatial wave functions multiplying $|u\rangle$ and $|d\rangle$ along each path are independent of the spin eigenstates $|j\rangle$. (That is, we can neglect clock breakup.) In addition, we assume that the magnetic field changes adiabatically over the interferometer paths such that the occupation of the states $|j\rangle$ does not change; only their phase changes with time according to $|j\rangle \to |j\rangle e^{-i\omega_jt}$.  The effect of the magnetic field difference between the paths is then to change the magnetic energies of the states along each path:  $\omega_j\to \alpha\omega_j$ for the upper path and $\omega_j\to \beta\omega_j$ for the lower path.  This change is equivalent to different proper times along paths $|u\rangle$ and $|d\rangle$. Then if $t$ is the time elapsed between $t_i$ to $t_f$, the overlap between the two clock states, which determines the visibility, is given by
\begin{equation}
 \langle u(t_f)|d(t_f)\rangle=\left\langle \sum_j c_je^{-i\alpha\omega_j t}|j\rangle \right|\left. \sum_k c_k e^{-i\beta\omega_j t}|k\rangle \right\rangle
=\sum_j |c_j|^2 e^{i(\alpha-\beta)\omega_j t}
\end{equation}
We see that the orthogonality of the clock states for the two arms is independent of the initial relative phases of the spin states. If the state is not pure, then a discussion of what is left of orthogonality or distiguishability is beyond our scope.

\bigskip

{\bf S4. Verifying the clock complementarity rule (Fig.\,4)}

Here we give details regarding the procedure yielding Fig.\,4. We emphasize that all three parameters of Eq. 3, namely $V$, $C$ and $D_I$, are measured independently. The procedures for measuring $C(T_R)$ and $D_I(T_G)$ are described in the next section. $V$ is simply measured
from the imaged interference pattern, by fitting the density profile to a Gaussian modulated by a sine function\,\cite{FGBS}.

\begin{figure}
\begin{center}
\includegraphics[width=12cm, height=10.cm]{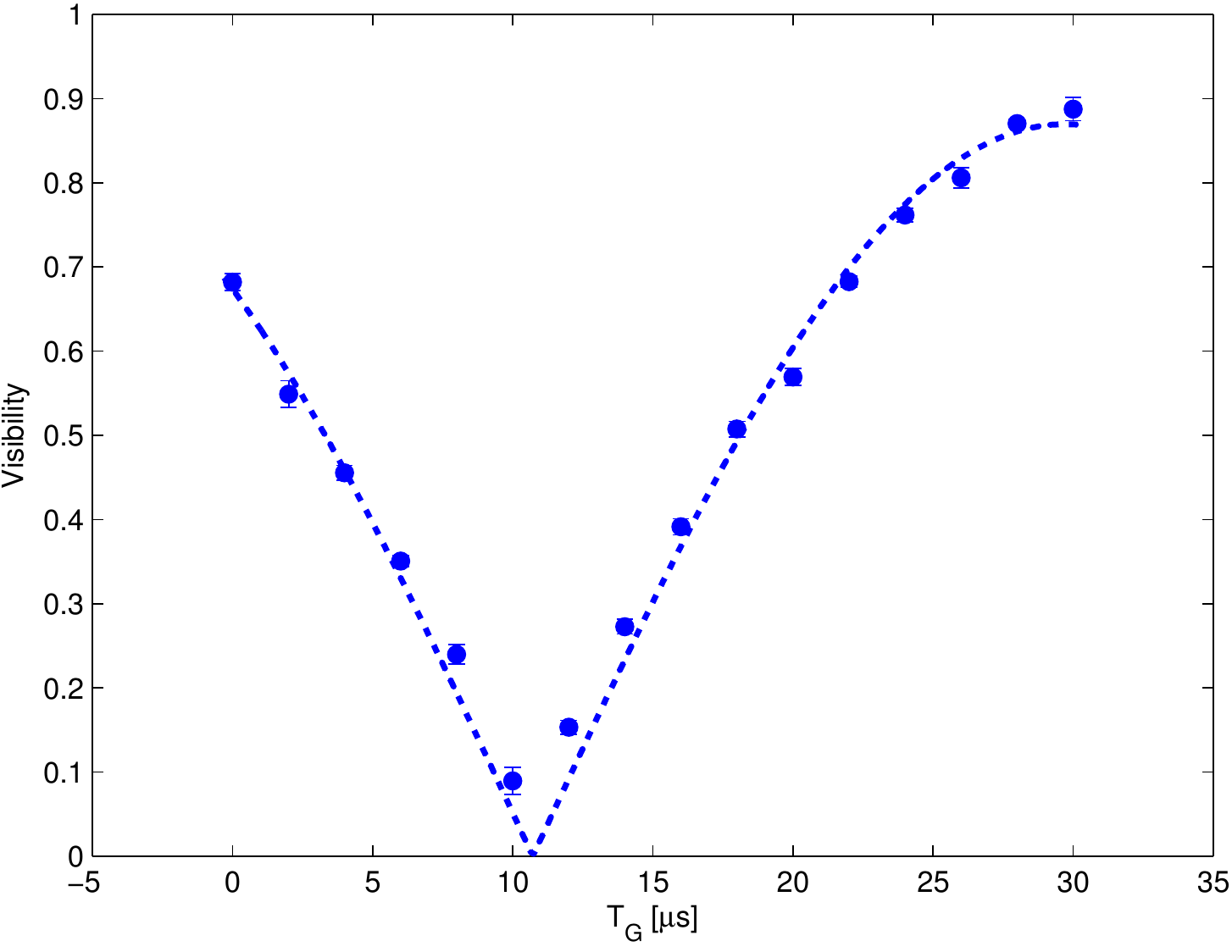}
\end{center}
\caption{The measured optimal-clock ($C=1$) interference visibility (without normalization) versus $T_G$, which induces distinguishability. The fitting function is $a\cdot|\cos[(\omega\cdot T_G+\phi_0)/2]|$, in which $a$ accounts for the irrelevant factors affecting visibility and $\phi_0$ is from the background gradient during the TOF stage after $T_G$.
}
\label{raw}
\end{figure}

Let us begin by presenting the high level of contrast achieved in the raw data for the visibility oscillations (Fig.\,\ref{raw}).

We then use the normalized visibility $V_{N}$ in order not to take into account irrelevant effects affecting visibility such as thermal background, imperfect focus, insufficient focal depth, imperfect overlap of the wave packets, etc. The procedure giving $V_{N}$ is simple: Each value of the visibility is an average of the single-shot visibility from several experimental cycles, and the error bars are the SEM (standard error of the mean) in this sub-sample. This average is normalized to the visibility of the single-state interferometer (i.e. without an initialization of a clock), with the latter ranging from 0.857$\pm$0.011 (SEM) to 0.891$\pm$0.013 (SEM) in different experimental runs.

\begin{figure}
\begin{center}
\includegraphics[width=12cm, height=10.cm]{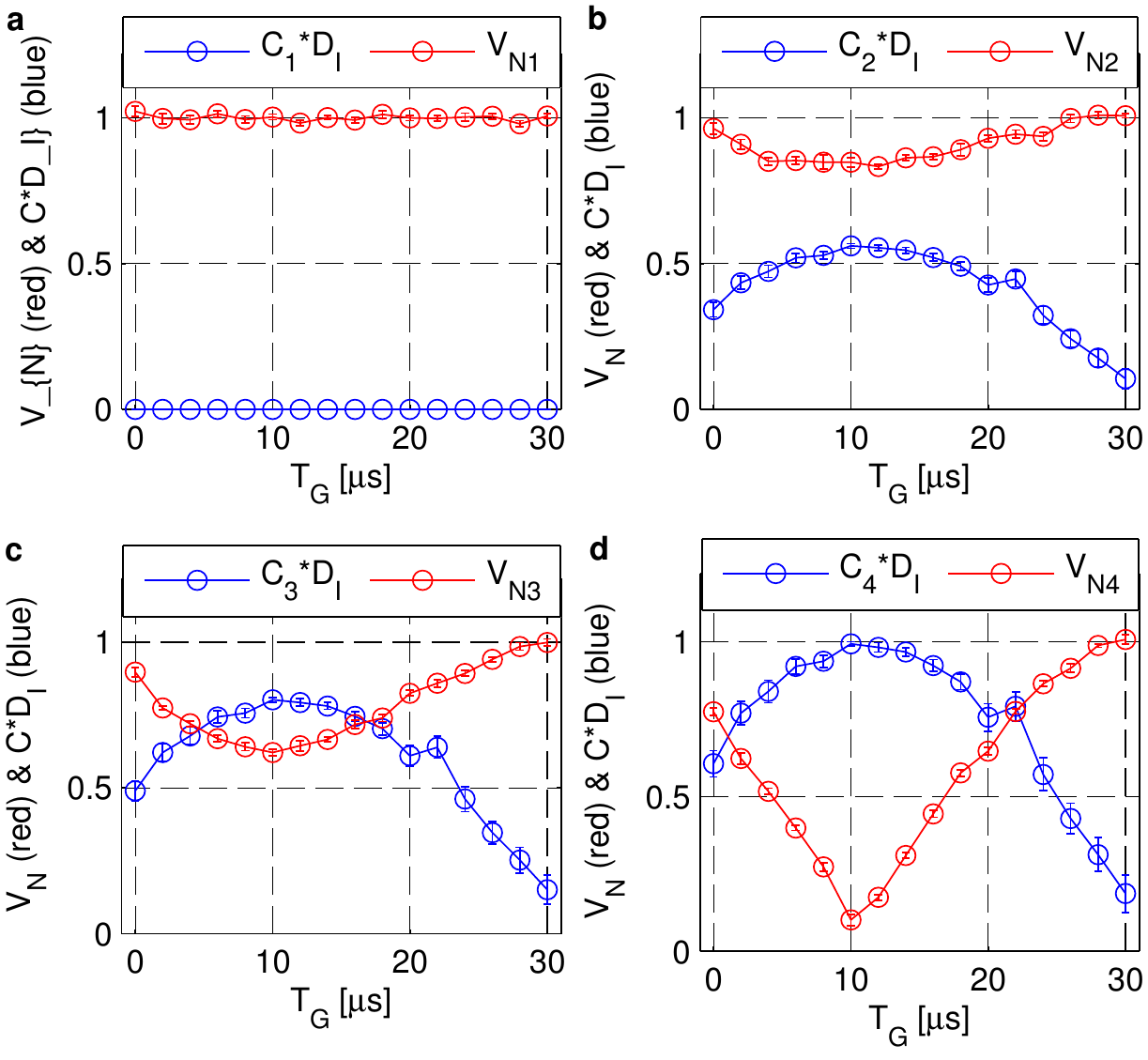}
\end{center}
\caption{({\bf a-d}), $V_{N}$ and $C\cdot D_I$ for 4 values of $C$ as in the main text: C=0, 0.56, 0.81, 1, for subplots ({\bf a-d}), respectively.
}
\label{fidvis}
\end{figure}

In Fig.\,\ref{fidvis} we present four subplots for $V_{N}$ and $C\cdot D_I$ when {\it C}=0, 0.56, 0.81, 1, as in the main text. When {\it C}=0 in Fig.\,\ref{fidvis}a, there is no projection on the equator and only one spin state is involved in the interference. Thus, the interferometric visibility $V_{N}$ is not influenced by $T_G$ (i.e. $D_I$). Instead, $T_G$ affects the phase of the interferometric fringes. As {\it C} starts to increase, the product of $C\cdot D_I$ starts to affect the visibility, causing a progressive drop in visibility until the maximal drop is reached when {\it C}=1. As presented in Fig.\,4 of the main text, the values shown in Fig.\,\ref{fidvis} give a total sum
$V^2_{N}+(C \cdot D_I)^2$ which equals 1 with high probability.

Finally, let us add that one type of clock imperfection that should be accounted for by the $C$ parameter in our specific realization of the clock interferometer is that of clock breakup due to the magnetic gradient simulating the GR redshift (as discussed in\,\cite{Us}). However, for our experimental parameters, we have evaluated this effect to be smaller than the experimental error bars ($\le 2\% $).

\bigskip

{\bf S5. Measuring $C$ and $D_I$ independently}

\begin{figure}
\begin{center}
\includegraphics[width=10cm, height=8.cm]{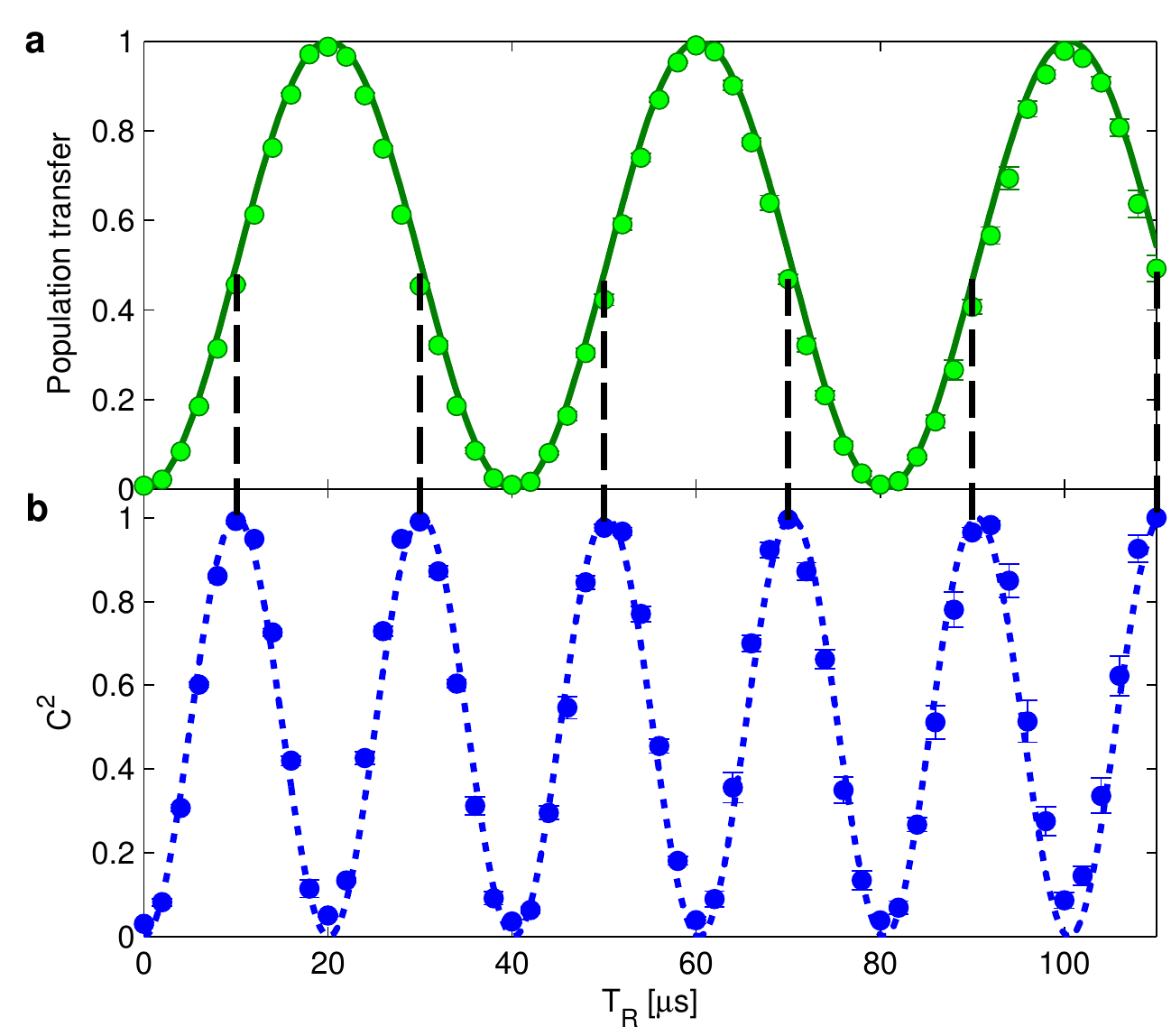}
\end{center}
\caption{{\bf a}, Measured population transfer $P$ vs. RF pulse duration $T_R$; {\bf b}, Values of $C^2$ vs. RF pulse duration $T_R$, where $C$ is defined as $2\sqrt{P(1-P)}$.
}
\label{fig4}
\end{figure}

We have previously described the corresponding relation between population transfer $P$ and $C$, such that $C=\sin\theta=2\sqrt{P(1-P)}$. To independently measure $C$ we thus need to measure the population transfer or the population balance. To measure the population transfer, we apply an RF pulse of duration $T_R$ at the same atom cloud position in which this pulse is applied in the real experiment. We do this while turning off all the additional subsequent pulses of the real experiment. We then use a strong Stern-Gerlach magnetic gradient to spatially split the two different spin states and measure how much population was transferred from the $m_F=2$ state to the $m_F=1$ state (Fig.\,\ref{fig4}a). $C(T_R)$ is then calculated using the measured result of $P$, as shown in Fig.\,\ref{fig4}b. From the figures, it can be seen that the period of $C$ is half the period of the population transfer, as expected.

\begin{figure}
\begin{center}
\includegraphics[width=12cm, height=10.cm]{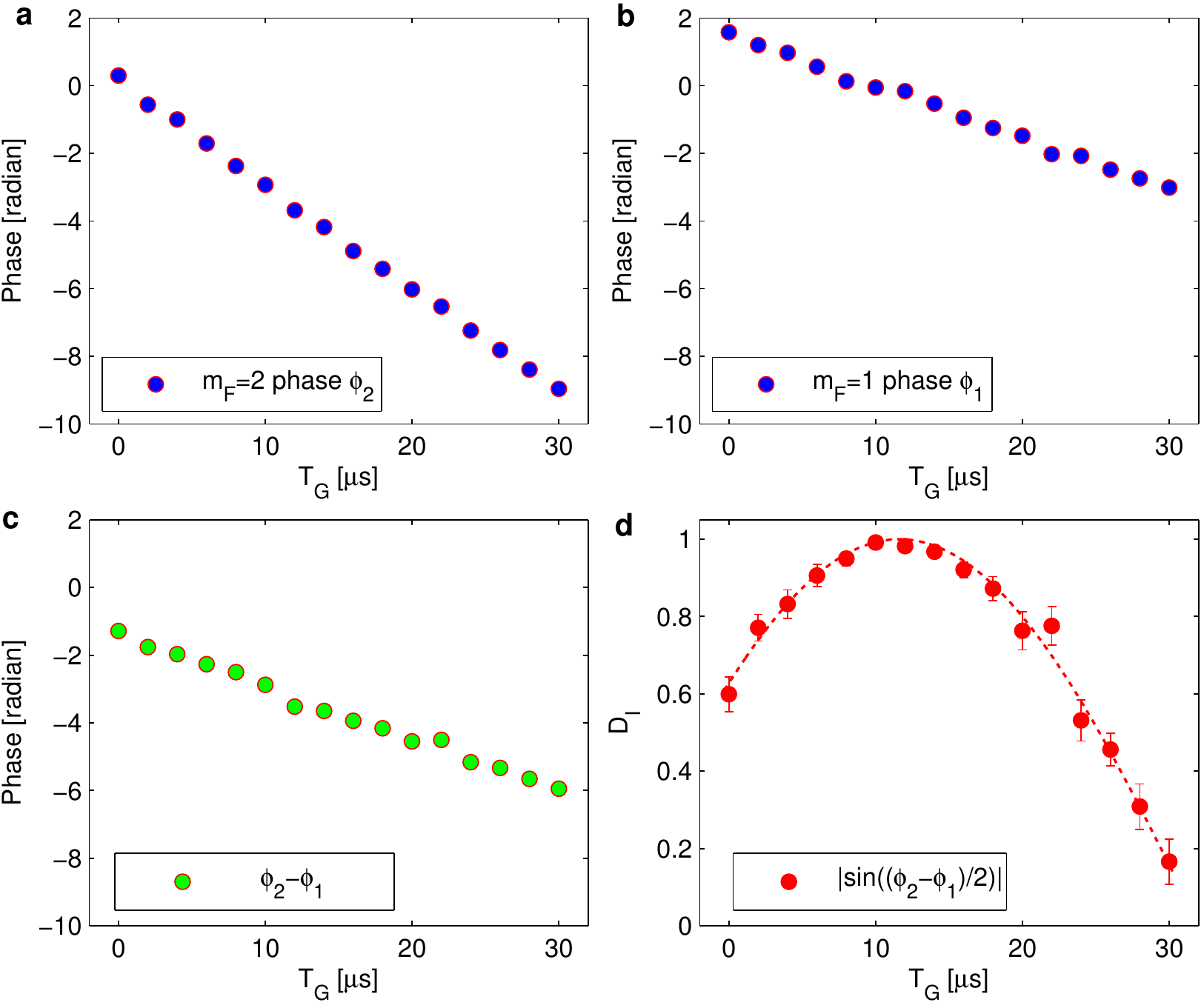}
\end{center}
\caption{Measuring $D_I$: {\bf a}, Measuring the $m_F=2$ single-state interferometer phase $\phi_2$. {\bf b}, Measuring the $m_F=1$ single-state interferometer phase $\phi_1$. {\bf c}, Calculating the phase difference $(\phi_2-\phi_1)$. {\bf d}, Calculating $D_I=|\sin[(\phi_2-\phi_1)/2]|$.
}
\label{fig5}
\end{figure}

$D_I(T_G)$ is measured, as shown in Fig.\,\ref{fig5}, in the following way: We do not prepare a clock but rather use an $m_F=2$ single-state interferometer to measure the phase difference between the two paths. We do the same for the $m_F=1$ interferometer. The difference between these two differential phases is equal to the phase difference between the clocks, from which $D_I$ is directly calculated.

\bigskip

{\bf S6. Applying the new rule to the Compton clock debate}

The clock complementarity relation may, for example, be applied to a single-state spatial interferometer, e.g. the so-called Compton clock \cite{Muller4}, for which $C=0$ (since the particle is not associated with any internal Hilbert space, or simply because $P=0$).  The parameter $C$, the ``clockness" or clock quality parameter, indicates how well a quantum system in a single position or along a single trajectory can measure time.
According to this definition of clock quality, $C=0$ does not correspond to a clock in the usual sense of the word; yet a single-state spatial interferometer can be sensitive to the relative passage of proper time along the interfering paths and thus constitute a clock (although -- as argued by Schleich and others -- not in the Kasevich-Chu configuration and without requiring the concept of the Compton frequency \cite{Wolfgang2,Wolfgang1,Wolfgang3,Wolfgang4,Peil}).
What is unique to $C>0$ clock interferometry is the reduced visibility of the interference pattern due to different clock readings along the paths rendering the paths distinguishable.

\bigskip
\goodbreak
\bigskip
\goodbreak
{\bf S7. Breaking the complementarity bound}
\nobreak

As an outlook, and following post-quantum speculations regarding ``less complementarity" \cite{Oppenheim}, and specifically the possibility of $V^2+D^2 > 1$ \cite{Zych}, let us briefly address such an hypothesis. For the latter to be true, at least one of the two assumptions used in constructing Eq. 3 must break down. This means that $V^2\equiv |\bra u |d\ket|^2$ and/or the rules of a scalar product in quantum formalism would be modified in some context.

Let us now assume, for example, that our fundamental assertion that $V^2\equiv |\bra u |d\ket|^2$ is wrong in the context of GR. In such a case, we can choose a perfect clock ($C=1$) and assume GR to give rise to $D_I=1$, while the visibility may still be high. For example, in the Bohmian picture there are trajectories \cite{Kocsis}. There is equal probability for a particle to be in any trajectory. In a double slit experiment, regions on the screen where few trajectories arrive form the dips in the interference pattern and regions where there is a high density of trajectories form the peaks. The ratio between high and low density of trajectories yields the visibility. As a naive outcome of a post-quantum model let us assume that in a time-independent setup these trajectories, and hence this ratio, are not affected by red shift, and consequently visibility is not affected even if a clock is traversing these trajectories and red shift brings the two clock wave packets to be orthogonal. As Bohmian mechanics has predictions equivalent to those of standard QM, the latter description obviously calls for an extension of the standard Bohmian mechanics.
In such a toy model let us fix $V=1$, while assuming $C^2$ and $D_I^2$ as in the main text. Indeed, in such a case it is easy to find parameters such that $V^2+(C \cdot D_I)^2 > 1$. Let us note that while to the best of our knowledge there is no general proof that $V^2 + D^2 > 1$ violates relativity (``no signalling"), some models that derive $V^2 +D^2 >1$ from stronger-than-quantum nonlocal correlations  \cite{Barrett} do violate relativity \cite{Rohrlich_Book}.
\bigskip

{\bf S8. No-clock interpretations}

In this section, we discuss visibility and point out that our predictions for visibility remain the same even though we can interpret them in two different, and complementary, ways.  In one interpretation, we regard atoms as two-level systems functioning as clocks; in the other interpretation, we consider only atomic spin eigenstates and their evolution.  For both cases, we first assume that the atomic states are coherent.  We then extend the discussion to include incoherent ensembles, i.e. mixtures, of atomic states.

In the case of coherent states (which we have discussed before \cite{Us}), the clock interferometer is in a superposition of two atomic spin states in two different ways.  On the one hand, we have a superposition of position eigenstates, i.e. atoms can take the upper path or the lower path of an interferometer; let us denote these states as $\psi_u (z)$ and $\psi_d (z)$, respectively.  In the interferometer, the position state is generally a superposition of the two paths.  On the other hand, we assume that every atom is in a superposition of internal states of a two-level atom, $\vert 1\rangle$ and $\vert 2 \rangle$, with respective energy levels $E_1$ and $E_2$.  As we know, such a superposition functions as an atomic clock via the relative dynamical phase $(E_2 - E_1)t/\hbar$, which causes (internal) precession proportional to the time $t$.

The interferometer superposes the atoms, for a time $T$, along the vertical axis with a separation $2z_0$. For simplicity let us assume that at the end of the duration $T$ the wave packets spread and overlap in a very short time $t\ll T$. This may be accomplished for example if during $T$ the wave packets are held in a tight potential. In analogy with two-slit interference, we can write these explicit one-dimensional wave functions for the atoms (after they have expanded freely) as
\begin{eqnarray}
\psi_u (z) &=&{{
e^{-(z-z_0)^2/[4\delta^2 +2i\hbar t/m]}}\over {(2\pi)^{1/4} \left[ \delta+i\hbar t/2m\delta \right]^{1/2}}}~~~,\cr
&{ }&\cr
\psi_d (z) &=&{{
e^{-(z+z_0)^2/[4\delta^2 +2i\hbar t/m]}}\over {(2\pi)^{1/4} \left[ \delta +i\hbar t/2m\delta \right]^{1/2}}}~~~,
\label{split}
\end{eqnarray}
where $z=0$ is the midpoint between the vertically separated wave packets, $\delta$ is the initial width of the wave packets, $m$ is the mass of the $^{87}$Rb atom and $t$ is the time since the wave packets began to expand freely.

The ``clock" superposition of internal states ``ticks" at a rate independent of $z$ if we treat time as a universal parameter as usual in quantum mechanics.  But to take into account the difference in proper times at $z=\pm z_0$, we apply the well-known formula from general relativity (e.g. \cite{grformula}): a time differential $\Delta T =Tg\Delta z/c^2$ accumulates between the wave packets, where $T$ is the time that the wave packets are separated in height by $\Delta z=2z_0$.  (Again for simplicity let us not include the wave-packet spreading time $t$ in $T$, because once the wave packets start to spread they are no longer completely separated by $\Delta z$.)  What is crucial is that the internal clock state becomes {\it entangled} with the position state.  We write
\begin{equation}
{1\over \sqrt{2}}\psi_u (z) [\cos{\theta\over 2} \vert 1\rangle + \sin {\theta\over 2} e^{i\phi+i\Delta \phi/2}\vert 2\rangle ]+
{1\over \sqrt{2}} \psi_d (z) [\cos{\theta\over 2}\vert 1\rangle + \sin {\theta\over 2}e^{i\phi-i\Delta \phi/2}  \vert 2\rangle
]~~~,
\label{enta}
\end{equation}
where (in this section) we take the angles $\theta$ and $\phi$ to be arbitrary, indicating the initial preparation of the clock state. What is not arbitrary is the relative phase $\Delta \phi$, which entangles time and position; it is due to the time differential $\Delta T$ that accumulates over the time $T$ in which the wave packets are separated in height by $\Delta z$, and equals $\Delta \phi = (E_2 - E_1)(\Delta T)/\hbar =2 gT z_0 (E_2-E_1)/c^2\hbar$.

Now, proper time affects every instance of time dependence in the overall wave function; but, except for the clock states themselves, we assume that its only effect is to induce a relative phase between the upper and lower paths, with negligible effect on the visibility of the interference.  We therefore neglect the effects of proper time on $\psi_u (z)$ and $\psi_d (z)$ in  Eq. (\ref{enta}).  What we cannot neglect is the effect of proper time on the clock states, which are now entangled with the position states in Eq. (\ref{enta}) whenever $\Delta T\ne 0$. It is this entanglement of clock time with vertical position that affects the visibility of the interference pattern.  As an illustration, let us consider a perfect clock ($C=1$, hence $\sin (\theta/2) =1/\sqrt{2}= \cos (\theta/2) $), and compute $V$ as a function of $\Delta \phi$:
\begin{equation}
V= \left\vert  (\langle 1\vert +e^{-i\phi +i\Delta \phi/2}\langle 2\vert)\cdot  (\vert 1\rangle +e^{i \phi +i\Delta \phi/2} \vert 2\rangle) \right\vert/2 =\vert \cos (\Delta\phi /2) \vert~~~~.
\label{vvp}
\end{equation}
(See also Sec. S3.)  Thus $V=1$ for $\Delta \phi= 2n\pi$, while for $\Delta \phi = (2n+1)\pi$ the clock states are orthogonal and the interference pattern disappears.  This calculation makes explicit the trade-off between the visibility of the interference and the ``which path" information or ``distinguishability" $D$ arising from the clock, formalized as $D^2+V^2\leq 1$.

In the alternative ``no-clock" interpretation, we can separately collect terms in ${1\over{2}}\psi_u (z) [\vert 1\rangle + e^{i\phi+i\Delta \phi/2}\vert 2\rangle ]+ {1\over {2}} \psi_d (z) [\vert 1\rangle + e^{i\phi-i\Delta \phi/2}  \vert 2\rangle ]$ corresponding to $\vert 1\rangle$ and $\vert 2 \rangle$ and calculate their separate interference patterns.  We obtain a sum of {\it two} interference patterns, proportional to
\begin{equation}\label{phi}
\vert \psi_u (z) +\psi_d(z) \vert^2  + \vert e^{i\Delta \phi/2}\psi_u (z)  +e^{-i\Delta \phi/2} \psi_d(z) \vert^2 ~~~,
\end{equation}
in which the first of the superimposed interference patterns comes from the $\vert 1 \rangle$ state and the second from the $\vert 2\rangle$ state, with a relative phase between the interference patterns that yields perfect visibility when $\Delta \phi =0$ (up to additions of multiples of $2\pi$) and zero visibility when $\Delta \phi = \pi$ (up to additions of multiples of $2\pi$), in agreement with Eq.\,(\ref{vvp}) yet without any explicit reference to a clock.  However, any loss of visibility has precisely the same {\it physical} origin in the two calculations:  the relative shift of the interference patterns is due to the proper-time differential between the two paths, taken into account by $\Delta \phi$.

Each calculation, with its associated interpretation, may have advantages and disadvantages in different contexts; but here we point out three advantages of the clock interpretation.  First, there are innumerable physical realizations of a clock, and innumerable clock characteristics, such as accuracy and mass.  For example, a superposition of $N > 2$ orthogonal states could form an accurate clock, and an external (rather than internal) variable could also serve to measure time \cite{ar}.  For any possible clock there would be a ``no-clock" mathematical analysis analogous to the one above; for example, instead of using a superposition of $N$ orthogonal states one could resort to $N$ independent interference patterns that would add constructively to yield perfect visibility or destructively to produce a flat probability distribution.  What all these analyses might miss is the insight that the system analyzed is a clock (and therefore must measure proper time).

Second, a ``no-clock" analysis, which isolates each energy level, does not connect the loss of visibility to ``which path" information.  From $V=|\cos(\phi/2)|$ there is zero visibility when $\phi =\pi$; but this loss of visibility arises as the sum of two interference patterns, each with full visibility. Since neither interference pattern is consistent with ``which path" information, how could we have guessed that their sum $is$ consistent with ``which path" information? For $N$ interference patterns the calculation would become cumbersome, all for lack of physical intuition about the relation between visibility and complementarity, a founding concept in quantum theory.

Third, clocks naturally lead us to consider time, hence also proper time and its possible relevance to quantum interferometry.  True, the effect of proper time does not depend on whether a clock is there to measure it.  The relative phase $\Delta \phi$ is due to gravitational proper time (red shift), which itself is due to the difference in height of the two paths: the proper times along the paths are different, inducing a relative phase between the clock states.  But a clock helps clarify why (and when) to take proper time into account.

One may generalize the above result beyond the coherent case.  For a statistical mixture of energy eigenstates, Eq.\,(\ref{phi}) still holds, and one may use the same arguments used in favour of the ``clock" interpretation.

Finally, let us consider another possible criticism of our clock interpretation.  This criticism asserts that there is no justification for introducing proper time {\it ad hoc}, as we have done.  Instead, given the two internal states $\vert 1 \rangle$ and $\vert 2 \rangle$ and their energy difference, which we have denoted $E_2-E_1$, we can obtain our relative phase result via special relativity, without clock states.  Each of these states interferes with itself over the two paths of the interferometer, and the interference terms (which can be calculated e.g. by obtaining the quantum phase from the action) include the potential energy factor $mg\Delta z$ (where $\Delta z$ is, as before, the difference in height between the paths).  However, according to special relativity, one can regard the internal states $\vert 1 \rangle$ and $\vert 2 \rangle$ as having different masses, because they have different energies.  Namely, applying Einstein's formula connecting energy and mass, we find that the mass of the $\vert 2\rangle$ state should be greater than the mass of the $\vert 1 \rangle$ state by $\Delta m = (E_2-E_1)/c^2$.  Now, if the two states have different masses, then the potential energy term $mg\Delta z$ will not be the same for the two states, because $m$ is not the same.   Replacing $m$ by $\Delta m$ in the formula for potential energy $mg\Delta z$, we obtain $(\Delta m) g \Delta z$ as the difference in potential energy.  Then, multiplying the result by $T/\hbar$, we find the relative phase $\Delta \phi$ between the $\vert 1\rangle$ and $\vert 2\rangle$ interference patterns to be $\Delta \phi = gT(\Delta z) (E_2-E_1)/c^2 \hbar$, exactly as obtained above in our proper-time calculation. The visibility is thus
\begin{equation}
V=|\cos(\Delta \phi/2)|=|\cos(\Delta E g \Delta z T/ 2\hbar c^2)| =|\cos(\Delta E \Delta V T/ 2\hbar c^2)|,
\end{equation}
where $\Delta V\equiv g \Delta h$.
This expression for the visibility is exactly the same as Eq. 13 in \cite{Zych}, which is based on a proper time calculation.

Without entering into a discussion of the merits of this special-relativistic calculation, we note that it identifies the {\it gravitational} mass for calculating the potential energy (i.e. the mass $m$ in $mg\Delta z$) with the {\it inertial} mass derived from the energy term in the phase via Einstein's formula $m = E/c^2$; the inertial mass takes into account not only the atomic mass but also the masses associated with the different internal energies.  By identifying the gravitational mass with the inertial mass, the special-relativistic calculation implicitly invokes the equivalence principle; indeed, the formula for gravitational proper time \cite{grformula} follows directly from the equivalence principle.
Therefore this interpretation, as well, is based on general relativity.

In conclusion, we cannot avoid using the equivalence principle if we are to account for the influence of a gravitational potential.  And, since general relativity implies a role for proper time, it does not make sense to try to avoid proper time as a physical parameter here.  Given the role of proper time, a description based on self-interfering clocks provides the best physical understanding, as argued above.


\clearpage


\begin{thebibliography}{99}

\bibitem{Kasevich} Kovachy, T. {\it et al.} Quantum superposition at the half-metre scale. {\it Nature} {\bf 528,} 530-533 (2015).

\bibitem{Tower} Muntinga, H. {\it et al.} Interferometry with Bose-Einstein condensates in microgravity.  {\it Phys. Rev. Lett} {\bf 110,} 093602 (2013).
\bibitem{muller} Yu, C., Estey, B., Zhong, W., Parker, R. H. \& M\"uller, H.
Improved accuracy of atom interferometry using Bragg diffraction,
in {\it Proceedings of the Seventh Meeting on CPT and Lorentz Symmetry (CPT16),}
Indiana University, Bloomington (2016).

\bibitem{JunYe} Nicholson, T. L. {\it et al.} Systematic evaluation of an atomic clock at 2$\times 10^{-18}$ total uncertainty. {\it Nat. Commun.} {\bf 6,} 6896 (2015).

\bibitem{Zych} Zych, M., Costa, F., Pikovski, I. \& Brukner, {\v C}. Quantum interferometric visibility as a witness of general relativistic proper time. {\it Nat. Commun.} {\bf 2,} 505 (2011).

\bibitem{Bushev}  Bushev, P. A., Cole, J. H., Sholokhov, D., Kukharchyk, N., \& Zych, M., Single electron relativistic clock interferometer. {\it  New J. Phys.} {\bf 18}, 093050 (2016).

\bibitem{Us} Margalit, Y. {\it et al.} A self-interfering clock as a `` which path'' witness. {\it Science} {\bf 349,} 1205 (2015).
\bibitem{Bohr} Bohr, N. Das Quantenpostulat und die neuere Entwicklung der Atomistik. {\it Naturwissenschaften} {\bf 16,} 245 (1928).

\bibitem{Englert} Englert, B.-G. Fringe visibility and which-way information: an inequality. {\it Phys. Rev. Lett.}. {\bf 77,} 2154 (1996).
\bibitem{Greenberg} Greenberger, D. M. \& Yasin, A. Simulteneous wave and particle knowledge in a neutron interferometer. {\it Phys. Lett. A} {\bf 128,} 391 (1988).
\bibitem{Jaeger} Jaeger, G., Shimony, A. \& Vaidman, L. Two interferometric complementarities. {\it Phys. Rev. A} {\bf 51,} 54 (1995).
\bibitem{Mandel} Mandel, L. Coherence and indistinguishability. {\it Opt. Lett.} {\bf 16,} 1882 (1991).
\bibitem{Zurek} Wootters, W. K. \& Zurek, W. H. Complementarity in the double-slit experiment: quantum nonseparability and a quantitative statement of Bohr's principle. {\it Phys. Rev. D} {\bf 19,} 473 (1979).
\bibitem{Aspect} Jacques, V. {\it et al.} Delayed-choice test of quantum complementarity with interfering single photons. {\it Phys. Rev. Lett.} {\bf 100,} 220402 (2008).
\bibitem{Rempe} Durr, S., Nonn, T. \& Rempe, G. Fringe visibility and which-way information in an atom interferometer. {\it Phys. Rev. Lett.} {\bf 81,} 5705 (1998).
\bibitem{Rauch} Summhammer, J., Badurek, G., Rauch, H., Kischko, U. \& Zeilinger, A. Direct observation of fermion spin superposition by neutron interferometry. {\it Phys. Rev. A} {\bf 27,} 2523 (1983).
\bibitem{MandelExp} Zou, X. Y., Wang, L. J. \& Mandel, L. Induced coherence and indistinguishability in optical interference. {\it Phys. Rev. Lett.} {\bf 67,} 318 (1991).

\bibitem{Haroche2} Bertet, P. {\it et al.} A complementarity experiment with an interferometer at the quantum-classical boundary. {\it Nature} {\bf 411,} 166 (2001).
\bibitem{Electron} Buks, E., Schuster, R., Heiblum, M., Mahalu, D. \& Umansky, V.
Dephasing in electron interference by a `which-path' detector. {\it Nature} {\bf 391,} 871 (1998).
\bibitem{Pfau} Pfau, T., Spalter, S., Kurtsiefer, C., Ekstrom, C. R. \& Mlynek, J. Loss of spatial coherence by a single spontaneous emission. {\it Phys. Rev. Lett.} {\bf 73,} 1223 (1994).
\bibitem{Chapman} Chapman, M. S. {\it et al.} Photon scattering from atoms in an atom interferometer: coherence lost and regained. {\it Phys. Rev. Lett.} {\bf 75,} 3783 (1995).
\bibitem{Bergou} Englert, B.-G. \& Bergou, J. Quantitative quantum erasure. {\it Opt. Commun.} {\bf 179,} 337 (2000).
\bibitem{NCinternal} Banaszek, K., Horodechi, P., Karpinski, M. \& Radzewicz, C.
Quantum mechanical which-way experiment with an internal degree of freedom. {\it Nat. Commun.}
 {\bf 4,} 2594 (2013).

\bibitem{Wineland} Chou, C. W., Hume, D. B, Rosenband, T., Wineland, D. J. Optical clocks and relativity.  {\it Science} {\bf 329}, 1630 (2010).

\bibitem{Campbell} Campbell, S. L., et al., A Fermi-degenerate three-dimensional
optical lattice clock.  {\it Science} {\bf 358}, 90 (2017).

\bibitem{Huntemann} Huntemann, N., Sanner, C., Lipphardt, B., Tamm, C. \& Peik, E. Single-ion atomic clock with 3$\times 10^{-18}$ systematic uncertainty.  {\it Phys. Rev. Lett.} {\bf 116},  063001 (2016).

\bibitem{Chen} Chen, J.-S., Brewer, S. M., Chou, C. W., Wineland, D. J., Leibrandt, D. R. \& Hume, D. B., Sympathetic ground-state cooling and time-dilation shifts in an ${^{27}}$Al$^+$ optical clock.  {\it Phys. Rev. Lett.} {\bf 118},  053002 (2017).

\bibitem{Ludlow} Ludlow, A. D., Boyd, M. M., Ye, J., Peik, E. \& Schmidt, P. O.  Optical atomic clocks. {\it Rev. Mod. Phys.} {\bf 87}, 637 (2015).

\bibitem{brukner} Pikovski, I., Zych, M., Costa, F. \& Brukner, {\v C}.
Time dilation in quantum systems and decoherence. {\it New J. Phys.} {\bf 19}, 025011 (2017).

\bibitem{Wolfgang1} Greenberger, D. M., Schleich, W. P. \& Rasel, E. M.
Relativistic effects in atom and neutron interferometry and the differences between them.
{\it Phys. Rev. A} {\bf 86,} 063622 (2012).

\bibitem{Lugli} Lugli, M., graduate thesis, Pavia (2017), arXiv 1710.06504
[quant-ph].

\bibitem{BE} Bohr, N. Discussions with Einstein on epistemological problems
in atomic physics, in {\it Albert Einstein: Philosopher--Scientist,} ed. Paul A. Schilpp (New York:  Tudor Pub. Co.),
201-41 (1951).

\bibitem{AR} Aharonov, Y. \& Rohrlich, D.
{\it Quantum Paradoxes:  Quantum Theory for the Perplexed} (Weinheim: Wiley-VCH), 2005, Sect. 2.4.

\bibitem{feynman lectures} Feynman, R., Leighton, R. \& Sands, M. {\it The
Feynman Lectures on Physics}, Vol. {\bf I} (Reading, MA: Addison�Wesley Pub. Co.) 1963, Sect. 42-6.

\bibitem{Muller4} Lan, S.-Y. {\it et al.}
 A clock directly linking time to a particle's mass. {\it Science} {\bf 339,} 554 (2013).

\bibitem{CCT2} Wolf, P. {\it et al.}
Does an atom interferometer test the gravitational redshift at the Compton frequency?
{\it Class. Quantum Grav.} {\bf 28,} 145017 (2011).

\bibitem{Wolfgang2} Schleich, W. P., Greenberger, D. M. \& Rasel, E. M.
Redshift controversy in atom interferometry: representation dependence of the origin of phase shift.
 {\it Phys. Rev. Lett.} {\bf 110,} 010401 (2013).

\bibitem{Peil} Peil, S. \& Ekstrom, C. R.
Analysis of atom-interferometer clocks. {\it Phys. Rev. A} {\bf 89,} 014101 (2014).

\bibitem{FGBS} Machluf, S., Japha, Y. \& Folman, R. Coherent Stern-Gerlach momentum splitting on an atom chip. {\it Nat. Commun.} {\bf 4,} 2424 (2013).

\bibitem{Lovecchio2016} Lovecchio, C. {\it et al.} Optimal preparation of quantum states on an atom-chip device.  {\it Phys. Rev. A} {\bf 93,} 010304(R) 2016.

\bibitem{Petrovic2013} Petrovic, J., Herrera, I., Lombardi, P., Schaefer, F. \& Cataliotti, F. S.
A multi-state interferometer on an atom chip. {\it New J. Phys.} {\bf 15,} 043002 (2013).

\bibitem{Budker} Derek, F., Kimball, J., Sushkov, A. O. \& Budker, D.
Precessing ferromagnetic needle magnetometer. {\it Phys. Rev. Lett.} {\bf 116,} 190801 (2016).

\end{thebibliography}

\begin{thebibliography}{99}

\bibitem{BE} Bohr, N. Discussions with Einstein on epistemological problems in atomic physics,
in {\it Albert Einstein: Philosopher--Scientist,} ed. Paul A. Schilpp (New York:  Tudor Pub. Co.),
201-41\,(1951).

\bibitem{Sinha} Sinha, S. \& Samuel, J.  Quantum limit on time measurement in a gravitational field. {\it Class. Quantum Grav.} {\bf 32,} 015018\,(2015).

\bibitem{Kasevich} Kovachy, T. {\it et al.} Quantum superposition at the half-metre scale. {\it Nature} {\bf 528,} 530-533\,(2015).

\bibitem{Tower} Muntinga, H. {\it et al.} Interferometry with Bose-Einstein condensates in microgravity.  {\it Phys. Rev. Lett} {\bf 110,} 093602\,(2013).
\bibitem{muller} Yu, C., Estey, B., Zhong, W., Parker, R. H. \& M\"uller, H.
Improved accuracy of atom interferometry using bragg diffraction,
in {\it Proceedings of the Seventh Meeting on CPT and Lorentz Symmetry (CPT’16),}
Indiana University, Bloomington\,(2016).

\bibitem{JunYe} Nicholson, T. L. {\it et al.} Systematic evaluation of an atomic clock at 2$\times 10^{-18}$ total uncertainty. {\it Nat. Commun.} {\bf 6,} 6896\,(2015).

\bibitem{Zych} Zych, M., Costa, F., Pikovski, I. \& Brukner, C. Quantum interferometric visibility as a witness of general relativistic proper time. {\it Nat. Commun.} {\bf 2,} 505\,(2011).
\bibitem{Us} Margalit, Y. {\it et al.} A self-interfering clock as a `` which path'' witness. {\it Science} {\bf 349,} 1205\,(2015).
\bibitem{Bohr} Bohr, N. Das Quantenpostulat und die neuere Entwicklung der Atomistik. {\it Naturwissenschaften} {\bf 16,} 245\,(1928).

\bibitem{Englert} Englert, B.-G. Fringe visibility and which-way information: an inequality. {\it Phys. Rev. Lett.}. {\bf 77,} 2154\,(1996).
\bibitem{Greenberg} Greenberger, D. M. \& Yasin, A. Simulteneous wave and particle knowledge in a neutron interferometer. {\it Phys. Lett. A} {\bf 128,} 391\,(1988).
\bibitem{Jaeger} Jaeger, G., Shimony, A. \& Vaidman, L. Two interferometric complementarities. {\it Phys. Rev. A} {\bf 51,} 54\,(1995).
\bibitem{Mandel} Mandel, L. Coherence and indistinguishability. {\it Opt. Lett.} {\bf 16,} 1882\,(1991).
\bibitem{Zurek} Wootters, W. K. \& Zurek, W. H. Complementarity in the double-slit experiment: quantum nonseparability and a quantitative statement of Bohr's principle. {\it Phys. Rev. D} {\bf 19,} 473\,(1979).
\bibitem{Aspect} Jacques, V. {\it et al.} Delayed-choice test of quantum complementarity with interfering single photons. {\it Phys. Rev. Lett.} {\bf 100,} 220402\,(2008).
\bibitem{Rempe} Durr, S., Nonn, T. \& Rempe, G. Fringe visibility and which-way information in an atom interferometer. {\it Phys. Rev. Lett.} {\bf 81,} 5705\,(1998).
\bibitem{Rauch} Summhammer, J., Badurek, G., Rauch, H., Kischko, U. \& Zeilinger, A. Direct observation of fermion spin superposition by neutron interferometry. {\it Phys. Rev. A} {\bf 27,} 2523\,(1983).
\bibitem{MandelExp} Zou, X. Y., Wang, L. J. \& Mandel, L. Induced coherence and indistinguishability in optical interference. {\it Phys. Rev. Lett.} {\bf 67,} 318\,(1991).

\bibitem{Haroche2} Bertet, P. {\it et al.} A complementarity experiment with an interferometer at the quantum-classical boundary. {\it Nature} {\bf 411,} 166\,(2001).
\bibitem{Electron} Buks, E., Schuster, R., Heiblum, M., Mahalu, D. \& Umansky, V.
Dephasing in electron interference by a `which-path' detector. {\it Nature} {\bf 391,} 871\,(1998).
\bibitem{Pfau} Pfau, T., Spalter, S., Kurtsiefer, C., Ekstrom, C. R. \& Mlynek, J. Loss of spatial coherence by a single spontaneous emission. {\it Phys. Rev. Lett.} {\bf 73,} 1223\,(1994).
\bibitem{Chapman} Chapman, M. S. {\it et al.} Photon scattering from atoms in an atom interferometer: coherence lost and regained. {\it Phys. Rev. Lett.} {\bf 75,} 3783\,(1995).
\bibitem{Bergou} Englert, B.-G. \& Bergou, J. Quantitative quantum erasure. {\it Opt. Commun.} {\bf 179,} 337\,(2000).
\bibitem{NCinternal} Banaszek, K., Horodechi, P., Karpinski, M. \& Radzewicz, C.
Quantum mechanical which-way experiment with an internal degree of freedom. {\it Nat. Commun.}
 {\bf 4,} 2594\,(2013).

\bibitem{Muller4} Lan, S.-Y. {\it et al.}
 A Clock Directly Linking Time to a Particle’s Mass. {\it Science} {\bf 339,} 554\,(2013).

\bibitem{CCT2} Wolf, P. {\it et al.}
Does an atom interferometer test the gravitational redshift at the Compton frequency?
{\it Class. Quantum Grav.} {\bf 28,} 145017\,(2011).

\bibitem{Wolfgang2} Schleich, W. P., Greenberger, D. M. \& Rasel, E. M.
Redshift controversy in atom interferometry: representation dependence of the origin of phase shift.
 {\it Phys. Rev. Lett.} {\bf 110,} 010401\,(2013).

\bibitem{Peil} Peil, S. \& Ekstrom, C. R.
Analysis of atom-interferometer clocks. {\it Phys. Rev. A} {\bf 89,} 014101\,(2014).

\bibitem{brukner} Pikovski, I., Zych, M., Costa, F. \& Brukner, C.
Time dilation in quantum systems and decoherence. {\it New J. Phys.} {\bf 19,} 025011\,(2017).

\bibitem{Wolfgang1} Greenberger, D. M., Schleich, W. P. \& Rasel, E. M.
Relativistic effects in atom and neutron interferometry and the differences between them.
{\it Phys. Rev. A} {\bf 86,} 063622\,(2012).

\bibitem{Lovecchio2016} Lovecchio, C. {\it et al.} Optimal preparation of quantum states on an atom-chip device.  {\it Phys. Rev. A} {\bf 93,} 010304(R)\,2016.

\bibitem{Petrovic2013} Petrovic, J., Herrera, I., Lombardi, P., Schaefer, F. \& Cataliotti, F. S.
A multi-state interferometer on an atom chip. {\it New Journal of Physics} {\bf 15,} 043002\,(2013).

\bibitem{Budker} Derek, F., Kimball, J., Sushkov, A. O. \& Budker, D.
Precessing ferromagnetic needle magnetometer. {\it Phys. Rev. Lett.} {\bf 116,} 190801\,(2016).

\bibitem{FGBS} Machluf, S., Japha, Y. \& Folman, R. Coherent Stern–-Gerlach momentum splitting on an atom chip. {\it Nat. Commun.} {\bf 4,} 2424\,(2013).

\bibitem{Binary2016} Li, R. B., {\it et al.} Rabi-oscillation-induced π phase flip in an unbalanced Ramsey atom interferometer. {\it Phys. Rev. A} {\bf 94,} 033613 (2016).

\bibitem{Binary2009}  Takahashi, A., Imai, H., Numazaki, K. \& Morinaga, A. Phase shift of an adiabatic rotating magnetic field in Ramsey atom interferometry
for m=0 sodium-atom spin states. {\it Phys. Rev. A} {\bf 80,} 050102{\bf (R)} (2009).

\bibitem{Binary2007} Usami, K. \& Kozuma, M. Observation of a Topological and Parity-Dependent Phase of m=0 Spin States. {\it Phys. Rev. Lett.} {\bf 99,} 140404 (2007).




\bibitem{Wolfgang3} Schleich, W. P., Greenberger, D. M. \& Rasel, E. M.
A representation-free description of the Kasevich–Chu interferometer: a resolution of the redshift controversy.
{\it New. J. Phys.} {\bf 15,} 013007\,(2013).

\bibitem{Wolfgang4} Giese, E. {\it et al.}
The interface of gravity and quantum mechanics illuminated by Wigner phase space.
{\it Proceedings of the International School of Physics “Enrico Fermi” Course 188 “Atom Interferometry”,} edited by G. M. Tino and M. A. Kasevich (IOS, Amsterdam; SIF, Bologna)\,2014.

\bibitem{Oppenheim} Oppenheim, J. \& Wehner, S. The Uncertainty Principle Determines the Nonlocality of Quantum Mechanics. {\it Science} {\bf 330,} 1072\,(2010).

\bibitem{Kocsis} Kocsis, S. {\it et al.} Observing the average trajectories of single photons in a two-slit interferometer. {\it Science} {\bf 332,} 1170\,(2011).

\bibitem{Barrett} Barrett, J. Information processing in generalized probabilistic theories. {\it Phy. Rev. A} {\bf 75,} 032304\,(2007).

\bibitem{Rohrlich_Book} Rohrlich, D. PR-Box Correlations Have No Classical Limit. in {\it Quantum Theory: A Two-Time Success Story,} 205, Struppa, D.C. \& Tollaksen, J.M. (eds.), Springer-Verlag Italia (2014); arXiv:1407.8530.

\bibitem{grformula} See, for example, Eq. (2.14) in Aharonov, Y. \& Rohrlich, D. {\it Quantum Paradoxes:  Quantum Theory for the Perplexed} (Weinheim:  Wiley-VCH), 2005.

\bibitem{ar} An example of an external variable as a measure of time appears in Aharonov, Y. \&  Rohrlich, D. {\it op. cit.} Sect.\,8.5.

\end{thebibliography}
\end{document}